\documentclass[fleqn,usenatbib]{mnras}

\usepackage{newtxtext,newtxmath}

\usepackage[T1]{fontenc}
\usepackage{datatool}

\usepackage[dvipsnames]{xcolor}

\definecolor{FFcol}{HTML}{E899D2}
\definecolor{IGMcol}{HTML}{9C9BF0}
\definecolor{SScol}{HTML}{98E676}
\definecolor{SWcol}{HTML}{0EBD93}
\definecolor{SIcol}{HTML}{F69F3D}
\definecolor{AScol}{HTML}{F9DF24}

\newcommand{\igm}{\textcolor{IGMcol}{{\bf Smooth IGM}}}
\newcommand{\ff}{\textcolor{FFcol}{{\bf Fountain flow}}}
\newcommand{\sw}{\textcolor{SWcol}{{\bf Satellite wind}}}
\newcommand{\si}{\textcolor{SIcol}{{\bf Satellite ISM}}}
\newcommand{\sts}{\textcolor{SScol}{{\bf Stripped satellite}}}
\newcommand{\accs}{\textcolor{AScol}{{\bf Accreted star}}}

\usepackage{graphicx}	
\usepackage{amsmath}	
\usepackage{amssymb}	
\newcommand\textlcsc[1]{\textsc{\MakeLowercase{#1}}}

\title[Cosmological fountain flows]{Gas accretion and galactic fountain flows in the Auriga cosmological simulations: angular momentum and metal re-distribution}

\author[R. J. J. Grand et al.]{\parbox[t]{\textwidth}{
Robert J. J. Grand$^{1}$\thanks{E-mail: grand@mpa-garching.mpg.de}, Freeke van de Voort$^1$, Jolanta Zjupa$^2$, Francesca Fragkoudi$^1$, Facundo A. G\'{o}mez$^{3,4}$, Guinevere Kauffmann$^1$, Federico Marinacci$^5$, R\"{u}diger Pakmor$^1$, Volker Springel$^1$, Simon D. M. White$^1$}
\vspace{10pt}
\\ 
$^1$Max-Planck-Institut f\"{u}r Astrophysik, Karl-Schwarzschild-Str. 1, 85748 Garching, Germany\\
$^2$Laboratoire Lagrange, Universite C\'{o}te d'Azur, Observatoire de la C\'{o}te d'Azur, CNRS, Blvd de l'Observatoire, \\ CS 34229, F-06304 Nice cedex 4, France \\
$^3$Instituto de Investigaci{\'o}n Multidisciplinar en Ciencia yTecnolog{\'i}a, Universidad de La Serena, Ra{\'u}l Bitr{\'a}n 1305, La Serena, Chile\\
$^4$Departamento de F{\'i}sica y Astronom{\'i}a, Universidad de LaSerena, Av. Juan Cisternas 1200 N, La Serena, Chile\\
$^5$Department of Physics \& Astronomy, University of Bologna, via Gobetti 93/2, 40129 Bologna, Italy
}

\date{Accepted XXX. Received YYY; in original form ZZZ}

\pubyear{2019}

\begin{document}
\label{firstpage}
\pagerange{\pageref{firstpage}--\pageref{lastpage}}
\maketitle

\begin{abstract}
Using a set of 15 high-resolution magnetohydrodynamic cosmological simulations of Milky Way formation, we investigate the origin of the baryonic material found in stars at redshift zero. We find that roughly half of this material originates from subhalo/satellite systems and half is smoothly accreted from the Inter-Galactic Medium (IGM). About $90 \%$ of all material has been ejected and re-accreted in galactic winds at least once. The vast majority of smoothly accreted gas enters into a galactic fountain that extends to a median galactocentric distance of $\sim 20$ kpc with a median recycling timescale of $\sim 500$ Myr. We demonstrate that, in most cases, galactic fountains acquire angular momentum via mixing of low-angular momentum, wind-recycled gas with high-angular momentum gas in the Circum-Galactic Medium (CGM). Prograde mergers boost this activity by helping to align the disc and CGM rotation axes, whereas retrograde mergers cause the fountain to lose angular momentum. Fountain flows that promote angular momentum growth are conducive to smooth evolution on tracks quasi-parallel to the disc sequence of the stellar mass-specific angular momentum plane, whereas retrograde minor mergers, major mergers and bar-driven secular evolution move galaxies towards the bulge-sequence. Finally, we demonstrate that fountain flows act to flatten and narrow the radial metallicity gradient and metallicity dispersion of disc stars, respectively. Thus, the evolution of galactic fountains depends strongly on the cosmological merger history and is crucial for the chemo-dynamical evolution of Milky Way-sized disc galaxies.
\end{abstract}

\begin{keywords}
galaxies: evolution - galaxies: formation - galaxies: spiral - galaxies: structure
\end{keywords}



\section{Introduction}

Inflows and outflows of gas in and around galaxies have emerged as crucial phenomena in galaxy formation. Shaped by cosmological gas accretion and energetic feedback, they mediate the transfer of mass, metals, (angular) momentum and energy between large-scale gas reservoirs (the CGM and IGM) and the interstellar medium (ISM) of galaxies \citep[e.g.][and references therein]{V17}. They are therefore integral to our understanding of the formation and properties of the stars that constitute galaxies.

Signatures of galactic winds are ubiquitously observed in the Milky Way \citep{WvW97} and star-forming galaxies \citep{SSP03} in the local Universe \citep{HAM90} and at high redshift \citep{PSS01,VCB05}. Both theory \citep[e.g.][]{CC85,BKS91} and observations \citep[e.g.][]{GFR14,LCA17} point toward stellar feedback and Active Galactic Nuclei (AGN) as viable mechanisms to provide the power necessary for driving galactic scale winds. While AGN are capable of driving outflows at velocities of thousands of kilometers per second \citep{FSU19}, galactic winds driven by stellar feedback can have velocities that lie below the escape speed of galactic haloes, which suggests that much of this material will be re-accreted, or ``wind recycled'' \citep{ODK10} in a galactic fountain. This is consistent with our understanding that gas from the hot corona/CGM must be continuously accreted onto the central galaxy in order for galaxies to sustain extended star formation histories \citep{TGN10,SKW11,SEM14}. 

Galactic winds likely have significant impact on the evolution of the angular momentum and metallicity distributions in the disc and surrounding inner halo, especially if they are powerful enough to extend into and mix with the CGM \citep[e.g.][]{HES19}. For example, \citet{FMM13} and \citet{F17} argue that the cold gas clouds around the Milky Way are part of a fountain flow that mix with the hot corona, and instigate the condensation of gas with high specific angular momentum acquired from large-scale tidal torques \citep[e.g.][]{P69,FE80,MMW98} at early times. However, it is unclear whether these cold gas complexes \citep{WvW97} are part of a galactic fountain flow \citep[see][]{MBF10,MFN11,FMA15}, or have been freshly accreted for the first time. Indeed, gas accretion onto galaxies may have external origins, including: gas brought into the main halo by satellite galaxies \citep{RPK12,LSB18}; and/or fresh gas smoothly accreted from the IGM that condenses gradually onto the galaxy \citep{VSB11,NGP16}. Gas properties such as kinematics and chemical composition will depend heavily on its origin and how it subsequently mixes, and will determine the chemodynamic properties of newborn stars over cosmic time. It is therefore highly desirable to understand how gas is accreted onto galaxies and contributes to star formation. However, this is extremely challenging observationally.

Hydrodynamical cosmological simulations enable us to explicitly follow the formation of galaxies, bringing forth the possibility to track gas flows as galaxies evolve \citep[e.g.][]{CMF10,FDO14,KNB19,SNG19}. Only recently, however, have numerical simulations successfully managed to produce star-forming spiral disc galaxies with prominent, rotationally supported disc components that reproduce numerous scaling relations \citep[e.g.][]{BSG12c,AWN13,MPS14,WDS15,CAR16,GGM17}. The general picture now emerging from simulations is that the formation of discs is promoted because energetic feedback expels low angular momentum gas from the central regions that later re-accretes onto the galaxy, providing fuel for late-time star formation from gas with, on average, high angular momentum \citep{BGR11,BSG12b,UNO14,TRD15,ZFB16,CDG16,AK16,DGB17} that should be well-aligned with the disc \citep[e.g.][]{SNT12}. Together, these results highlight the doubly-important role of feedback to both curtail early star formation and move gas from low- to high-angular momentum \citep{SD15,AK15,GFH15,NO17}.

The most modern cosmological zoom-in simulations of the formation of Milky Way mass galaxies are able to reach resolutions with $\sim 10^4$ $\rm M_{\odot}$ per baryonic element \citep[e.g.][]{GC11,FNS16,GGM17,GHW19,ODM19}, enabling the study of CGM properties \citep[e.g.][]{HES19} and galactic winds in finer detail. \citet{TCM19} analyse the gas flow properties in the context of their thermodynamic phase in a suite of simulations, and find that for Milky Way-mass systems, galactic fountains with typical recycling timescales of $\sim 1$ Gyr do not tend to mix with the hot CGM, in contrast to other theoretical work \citep{PFB17}. \citet{AAF17} analyse an independent set of simulations, and find a much shorter recycling timescale ($\sim 200$ Myr), and moreover demonstrate for their three Milky Way-mass galaxies that almost half of the baryonic material that forms stars by redshift zero originates from satellites in the form of ``inter-galactic wind transfer''.  Aside from the novel implication that galactic fountain flows may be a sub-dominant form of galactic gas accretion, it follows also that this wind material may influence the angular momentum content and orientation of the inner CGM with respect to the disc, which may itself be torqued into a tumble by mechanisms that arise from satellite interactions \citep{GWG17,GGM17b} and the accretion of massive subhaloes \citep{MGG16,MGG19}. Thus, the prevalence and evolutionary connection of galactic fountain flows to the growth of discs in the face of this profusion of galaxy formation processes over cosmic time is unclear.

The nature and evolutionary effects of simulated baryon cycles likely depend heavily on the galaxy formation model, numerical implementation thereof, and the codes employed in numerical simulations \citep[see for example][for a study of the Illustris and IllustrisTNG simulations]{KNB19}. In this contribution, we extend the Auriga simulation suite \citep[][described in Section~\ref{sec:sim}]{GGM17} and investigate the accretion history and origin of gas in a set of cosmological zoom-in simulations of Milky Way mass galaxies. Importantly, these simulations reproduce a range of properties expected for Milky Way mass galaxies: they are disc-dominated star-forming galaxies with flat rotation curves and reproduce a range of observed scaling relations such as the Tully-Fisher relation \citep{GGM17}, the size-mass relation of HI gas discs \citep{MGP16} and the mass function of Milky Way satellites \citep{SGG17}. This provides us with a solid basis on which to investigate the mechanisms and conditions for galactic fountain flows to help grow discs across cosmic time. We describe our simulations in Section~\ref{sec:sim}.

In what follows, we use a tracer particle analysis \citep{DGB17} to track the history of baryonic elements found in star particles at redshift zero (described in Section~\ref{sec:sim}), and classify the type of accretion based on a set of criteria described in Section~\ref{sec:class}. In Section~\ref{sec:results}, we show that, in general, nearly all of the stars formed by redshift zero contain previously wind-recycled material, to which galactic fountain flows contribute significantly. We study the wind-recycling times and radial extent associated with fountain flows, and confirm that in most cases they generally act to increase the angular momentum of gas and disc stars by mixing with high angular momentum corona material. We show that the rate of specific angular momentum growth depends on the disc-CGM alignment and the merger history, with implications for their evolution in the specific angular momentum-stellar mass relation \citep[][]{FR18} and metal distribution of the disc. We discuss some convergence and resolution concerns in Section~\ref{sec:conv} and summarise our results in Section~\ref{sec:conc}.

\begin{figure*}
\centering
\includegraphics[scale=1.65,trim={0 0.1cm 0 0},clip]{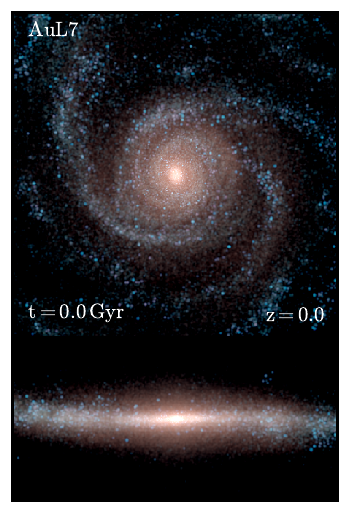}
\includegraphics[scale=0.8,trim={0 0 0 0},clip]{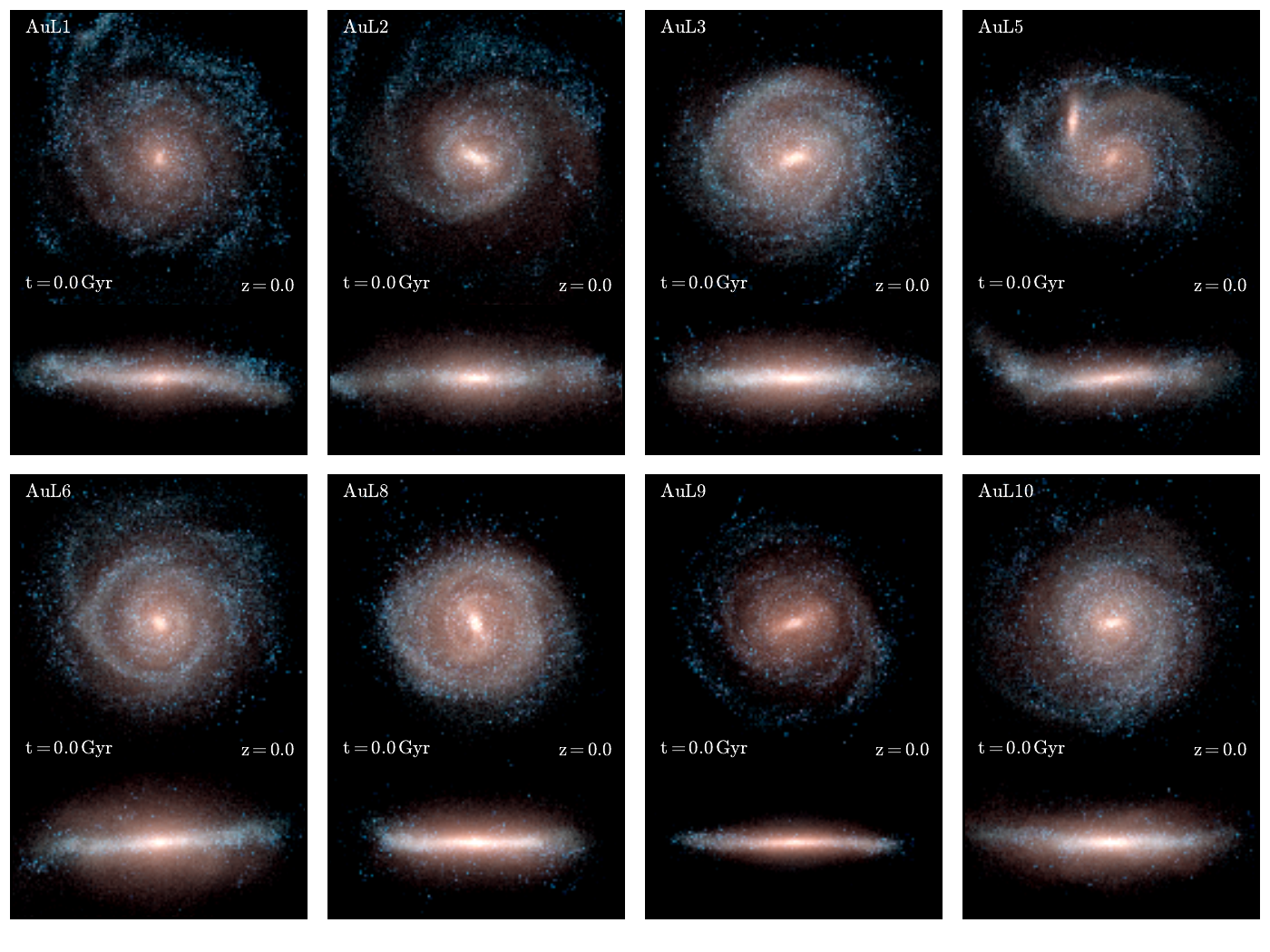}

\caption{Face-on and edge-on projected stellar densities for
  each of the new ``low-mass'' simulations. The images are a projection of the $K$-, $B$- and
  $U$-band luminosity of stars, shown by the red, green and blue
  colour channels, in logarithmic intervals, respectively. Younger
  (older) star particles are therefore represented by bluer (redder)
  colours. The box side-length is $50\times 50$ kpc for face-on images, and $50\times 25$ kpc for edge-on images. Movies and images are available at
  \href{https://wwwmpa.mpa-garching.mpg.de/auriga}{\url{https://wwwmpa.mpa-garching.mpg.de/auriga}}.} 
\label{fproj}
\end{figure*}

\section{Simulations \& Methods}
\label{sec:sim}
\subsection{Simulation Overview}

We study a total of 15 cosmological magneto-hydrodynamical zoom simulations. Six\footnote{The original set of Auriga simulations did not include tracer particles, therefore we include only a subset of simulations re-run with the tracer particles.} of these are taken from the original Auriga simulation suite \citep{GGM17}, which simulate the formation of star-forming disc galaxies in the halo mass\footnote{Defined to be the mass inside a sphere in which the mean matter density is 200 times the critical density, $\rho _{\rm crit} = 3H^2(z)/(8 \pi G)$.} range $1 < M_{200} / 10^{12}$ $\rm M_{\odot} < 2$. The other nine simulations presented here span the slightly lower halo mass range of  $0.5 < M_{200} / 10^{12}$ $\rm M_{\odot} < 1$, thus extending the Auriga suite to cover the entire possible mass range of the Milky Way \citep[see e.g.][for recent measurements]{CCD19,DFB19,GDW19}. We will hereafter refer to these nine simulations as ``low mass'', with subscripts of the form LN, where N denotes the halo number, to distinguish them from the original Auriga simulations. All simulations begin at $z=128$ with cosmological parameters:  $\Omega _m = 0.307$, $\Omega _b = 0.048$, $\Omega _{\Lambda} = 0.693$ and a Hubble constant of $H_0 = 100 h$ km s$^{-1}$ Mpc$^{-1}$, where $h = 0.6777$, taken from \citet{PC13}. Dark matter particles have a mass of $\sim 4 \times 10^{5}$ $\rm M_{\odot}$, and the baryonic mass resolution is $\sim 5 \times 10^{4}$ $\rm M_{\odot}$. The physical softening of collisionless particles grows with time (corresponding to a fixed comoving softening length of 500 pc $h^{-1}$) until a maximum physical softening length of 375 pc is reached, which are reasonable choices for this mass resolution \citep{PNJ03}. The physical softening value for the gas cells is scaled by the gas cell radius (assuming a spherical cell shape given the volume), with a minimum softening set to that of the collisionless particles.

The simulations are performed with the magneto-hydrodynamic code \textlcsc{AREPO} \citep{Sp10}, with a galaxy formation model  that includes: primordial and metal line cooling, a prescription for a uniform background UV field for reionization (completed at $z=6$), a subgrid model for star formation that activates for gas densities larger than $0.11$ atoms $\rm cm^{-3}$ \citep{SH03}, magnetic fields \citep{PMS14,PGG17,PGP18}, gas accretion onto black holes and energetic feedback from AGN and supernovae type II (SNII) \citep[see][for more details]{VGS13,MPS14,GGM17}. Each star particle represents a single stellar population with a given mass, age and metallicity. Mass loss and metal enrichment from type Ia supernovae (SNIa) and Asymptotic Giant Branch (AGB) stars are modelled by calculating at each time step the mass moving off the main sequence for each star particle according to a delay time distribution. The mass and metals are then distributed among nearby gas cells with a top-hat kernel. We track a total of 9 elements: H, He, C, O, N, Ne, Mg, Si and Fe. 

SNII feedback is modelled with a phenomenological wind model, the main parameters of which are the energy available per SNII per unit mass, $\epsilon _{\rm SN}$, and the wind velocity, $v_w$. We set the wind velocity to scale with the 1-dimensional velocity dispersion of local dark matter particles, $\sigma$, according to 

\begin{equation}
v_w = \kappa _{\rm kin} \sigma,
\end{equation}
where $\kappa _{\rm kin} = 3.46$ \citep{PS13,MPS14}. This scaling was shown in the simulations of \citet{OFJ10} to reproduce the observed satellite luminosity function and match the luminosity-metallicity relation of Local Group satellites. The SNII energy per unit mass is given by

\begin{equation}
\begin{split}
\epsilon _{\rm SN} &= \frac{\eta _w v_w^2}{2} + \frac{3 \eta _w \kappa _{\rm th} \sigma ^2}{2} \\ &= \frac{\eta _w v_w ^2}{2} \Bigg(1 + \frac{3 \kappa _{\rm th}}{\kappa _{\rm kin}^2} \Bigg),
\end{split}
\end{equation}
where the kinetic and thermal energy factors, $\kappa _{\rm kin}$ and $\kappa _{\rm th}$, are set such that the wind carries thermal and kinetic energy in equal parts. This expression thus yields the mass loading, $\eta _w$, defined as the ratio between the wind mass flux and the star formation rate. 

In our model, the probability for the $i$-th ISM gas cell to model star/wind formation in a given timestep, $\Delta t$, is given by

\begin{equation}
    p = \frac{M_i}{M_*} \Big( 1 - e^{-(1+\eta _w )\Delta t / t_{\rm SF}} \Big),
    \label{eqp1}
\end{equation}
where $t_{\rm SF}$ is the star formation timescale, and $M_i$ and $M_*$ are the masses of the gas cell and target star/wind particle mass, respectively. For cells that pass this random decision, we make a binary choice to treat either SNII wind formation (under the instantaneous approximation) or normal star formation according to 

\begin{equation}
p_{w,\rm SF} =
\begin{cases}
\eta _w / (1 + \eta _w) & \quad \text{for SNII winds}, \\
1 / (1 + \eta _w) & \quad \text{for star formation}.
\end{cases}
\label{eqp2}
\end{equation}
This guarantees that the correct amount of star formation and SNII winds are treated at each timestep. 

In practice, SNII winds are modelled by ``wind particles'', which are immediately launched in an isotropically random direction upon creation. They carry away the mass and $1-\eta _{\rm metal} = 0.4$ times the metallicity of the gas cell (the remaining metals are left behind in the gas at the launch site)\footnote{We note that the choice of $\eta _{\rm metal} = 0.6$ is made to reproduce the mass metallicity relation in galaxies in large cosmological box simulations \citep{VGS13}.}. They temporarily hydrodynamically de-couple from the two-phase subgrid ISM model \citep{SH03}, and recouple to lower density gas ($5\%$ of star formation density threshold) outside the ISM through a non-local momentum injection. Thus, wind particles in essence model the emergence of winds from the star-forming ISM. It is important to note that although this model does not fully capture the entrainment of dense ISM gas in winds, it does drive galactic-scale winds after the wind particles hydrodynamically re-couple to the gas outside the multi-phase ISM \citep{NPS19}.

\begin{figure}
\centering
\includegraphics[scale=1.5,trim={0 0 0 0.8cm},clip]{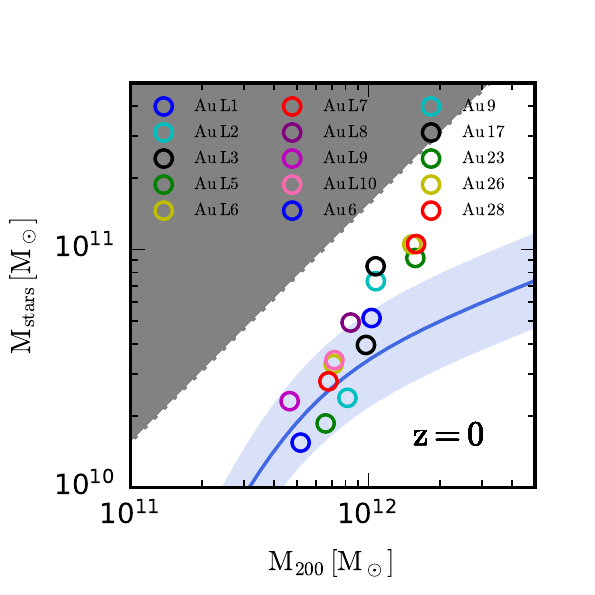}\\
    \caption{The stellar mass-halo mass relation at redshift zero for the simulations studied in this paper. The blue curve and shaded region indicate the abundance matching relation and scatter from \citet{MNW13}. The shaded grey region marks stellar masses above the universal baryon fraction. Galaxies below a halo mass of $\sim 10^{12}$ $\rm M_{\odot}$ lie on the abundance matching relation, whereas our sample of original Auriga galaxies above this mass lie above the relation, in good agreement with recent observations of nearby massive spiral galaxies \citep{PFM19}.}
    \label{am}
\end{figure}

In Fig.~\ref{fproj}, we show face-on and edge-on false-colour projections of the stellar light at redshift zero for each of the nine new simulations; the 6 from the original Auriga suite are shown in \citet{GGM17}. As in that work, the simulated galaxies possess radially extended, vertically thin spiral discs that surround an older, redder bulge/bar component. Fig.~\ref{am} shows stellar mass as a function of halo mass and demonstrates that the stellar masses of these haloes are in good agreement with abundance matching curves \citep[e.g.][]{MNW13}, with the exception of the 5 most massive galaxies which lie slightly above the relation - a trend that is remarkably similar to the stellar mass-halo mass relation inferred from HI rotation curves of nearby massive spiral galaxies \citep{PFM19}. This fascinating development will be followed up in a future study.

\subsection{Simulation data and analysis}

\begin{figure*}
\centering
\includegraphics[scale=0.5,trim={0 0.5cm 0 0.8cm},clip]{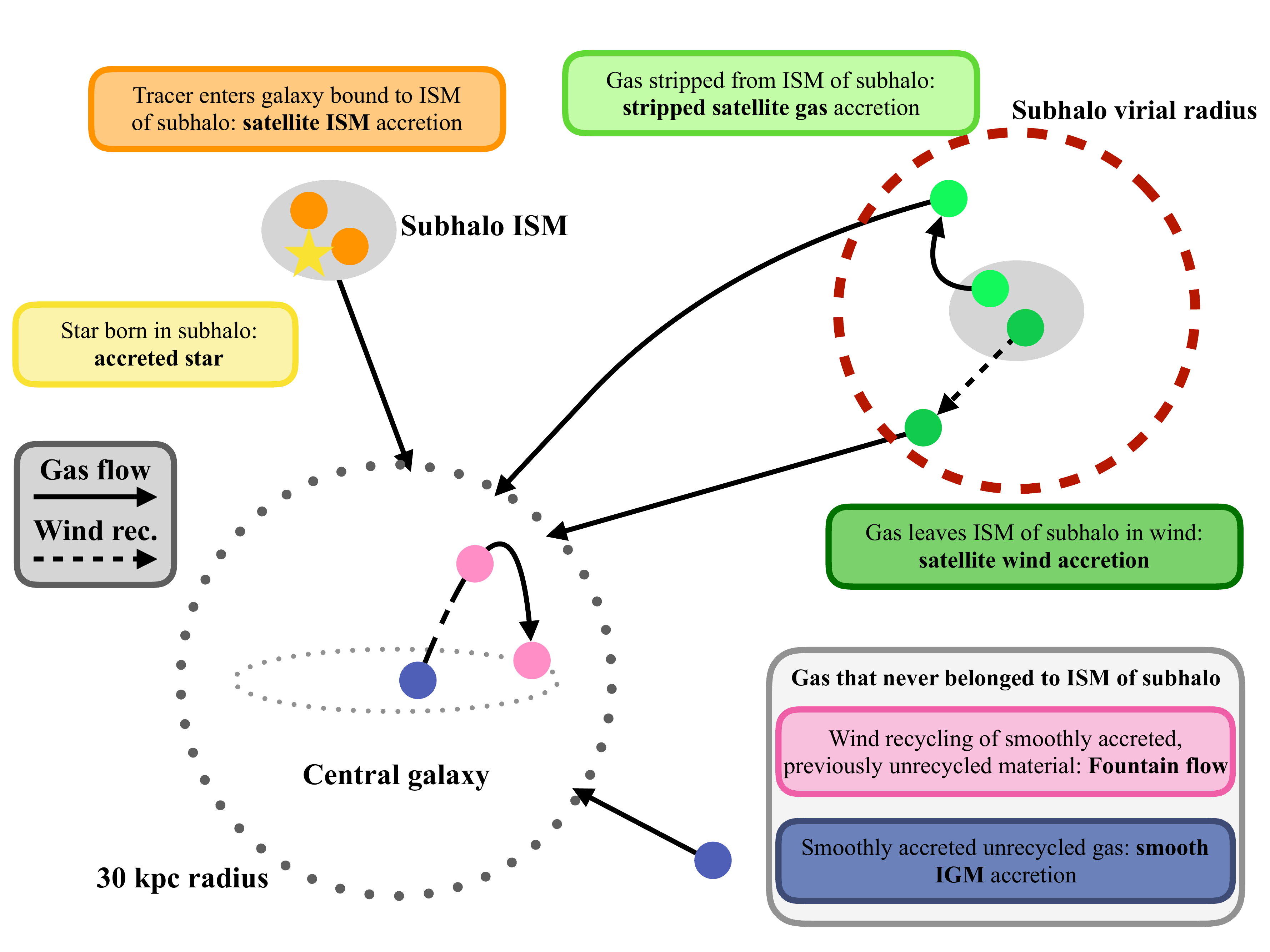}\\
    \caption{Illustration of the classification scheme for the accretion sources considered in this paper (see Section~\ref{sec:class} for details). Gas elements (stars) are depicted as solid circles (stars), coloured according to their classification. Trajectories of hydrodynamical flows are indicated by solid lines, whereas those driven by wind-recycling events are indicated by dashed lines. The 30 kpc radius of the central galaxy defines the accretion crossing point. The ISM of galaxies is indicated by the grey shaded regions.}
    \label{flowd}
\end{figure*}

Each simulation produces 252 snapshots (compared to 128 in the original Auriga simulations) that are spaced approximately 60 Myr apart. In addition to the raw particle/cell data, each snapshot contains halo catalogues returned by an on-the-fly structure finder \textlcsc{SUBFIND} \citep{SWT01} that identifies gravitationally bound subhaloes and central haloes within groups identified by a friends of friends (\textlcsc{FOF}) algorithm \citep{DEF85}. We used a modified version of the LHaloTree algorithm to construct merger trees in post-processing \citep[see][for details]{SGG17}, which links the merger history of all subhaloes between snapshots and enables us to identify and follow the same objects in time. We use these tools to assign gas cells and particles to individual subhaloes at each snapshot. We use the centre of mass of the central halo to calculate the distance of each particle/cell from the central galaxy at each snapshot. For the purposes of this analysis, we rotate the data in each snapshot to align the $Z$-coordinate with the principal axis of the stellar angular momentum inside 10 kpc, as described in \citet{GGM17}.

\subsubsection{Tracer particle analysis}
\label{tpa}

These simulations include Lagrangian ``tracer particles'' \citep[see][]{GVN13,DGB17}, which enable us to track the evolutionary history of gas cells of interest. Because the quasi-Lagrangian nature of the \textlcsc{AREPO} moving-mesh code entails the advection of gas across cell faces, a gas cell cannot be considered to contain the same gas throughout its evolution, as is commonly done in particle based methods such as Smoothed Particle Hydrodynamics \citep[as in][for example]{GKC11}. In general, a fraction of the gas contained in a single Voronoi cell may be advected to any adjacent cell in a single time-step, depending on the local fluid dynamics and cell geometry. This feature of the code means that it is appropriate to use a Monte-Carlo random sampling of all possible fluxes across each face of a Voronoi cell, with sampling probabilities proportional to the flux across each face. At the beginning of each simulation, each high-resolution gas cell (of equal mass) is assigned 1 tracer particle. These particles move across a cell face with a probability given by the ratio of the outward-moving mass flux across a face and the mass of the cell. Usually, the highest probability is for a tracer particle to remain in the same cell, because the cells follow the bulk gas flow to minimize advection.

Tracer particles that reside in gas that undergoes star formation become locked inside star particles. As stellar evolution proceeds, tracers may re-enter surrounding gas through mass loss from AGB stars and/or SNeIa. The probability for this to occur at a given timestep is rather low, given the relatively small mass fraction lost from the star particle. We have confirmed that the number of mass transfer events from star particles to the gas is subdominant with respect to SNII winds, and do not consider them further in this paper.

We proceed to track the history of the tracer particles residing in these stars back in time, and determine their position and angular momentum values at each snapshot. We also identify, at each snapshot, whether or not a tracer belongs to the main galaxy or a subhalo, and whether a tracer resides in a star-forming gas cell, non-star-forming gas cell, star particle or a wind particle. Crucially, the simulation output records the precise times and positions at which a tracer particle has entered a wind phase, therefore the wind-recycling times, defined to be the time between successive wind recycling events, are known exactly for all tracer particles and are independent of snapshot spacing/cadence.

\begin{figure*}
\centering
\includegraphics[scale=1.05,trim={2.cm 0.5cm 0 1.5cm},clip]{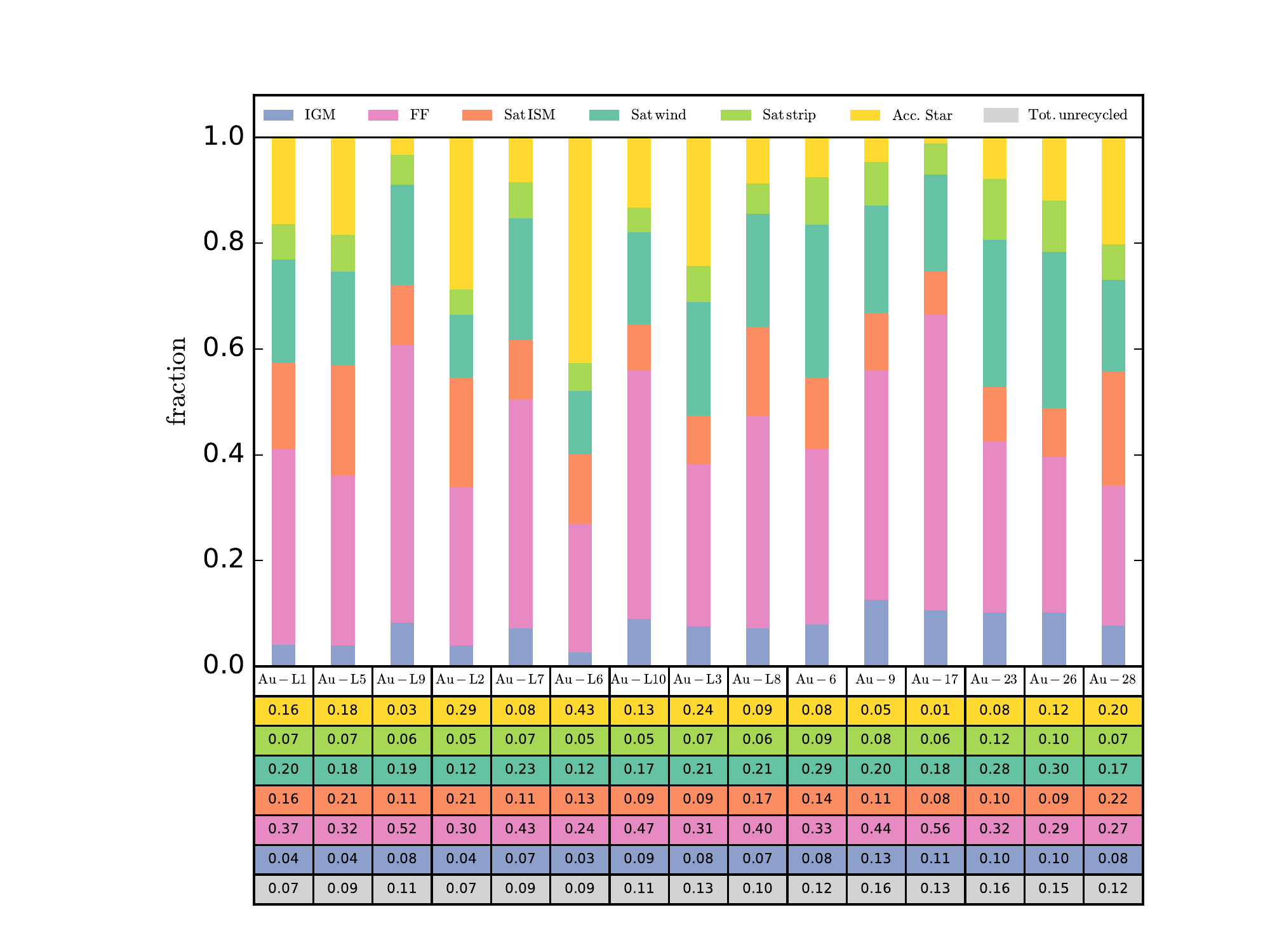}\\
    \caption{The mass fraction of accretion sources that make up star particles within a 30 kpc radius of the central galaxy at redshift zero, for each simulation (ordered by increasing stellar mass from left to right). The sources are: \igm, \ff, \si, \sw, \sts, and \accs. The bottom row of the table lists the total fraction of unrecycled material, which includes \igm { by} construction, but also contributions from \si, \sts { and} \accs, which each comprise recycled and unrecycled material.}
    \label{forig}
\end{figure*}

\section{Classification of baryonic material origin}
\label{sec:class}

Before describing the classification procedure, we note that the terms ``recycling'' and ``wind-recycling'' are used interchangeably throughout this paper, and are taken to refer to the winds generated by SNeII. In the technical framework of our simulations, this specifically refers to the generation of wind particles, as described in Section~\ref{sec:sim}. 

In the following, we identify the tracer particles that reside in star particles at $z=0$\footnote{Owing to the presence of a satellite within 30 kpc at $z=0$ in halo Au L5, we use the snapshot at $t_{\rm lookback}=1.5$ Gyr as the end-point to identify tracers in star particles.}, and track their evolutionary history to define six modes of accretion onto the central galaxy. These modes are: i) accretion of unrecycled gas from the IGM (\textcolor{IGMcol}{{\bf Smooth IGM}}); ii) gas that is wind recycled at least once after first accretion (but unrecycled before first accretion - \textcolor{FFcol}{{\bf Fountain flow}}); the accretion of satellite gas in the form of iii) direct \textcolor{SIcol}{{\bf Satellite ISM}} accretion; iv) \textcolor{SWcol}{{\bf Satellite wind}} transfer; v) \textcolor{SScol}{{\bf Stripped satellite}} ISM gas; and vi) \textcolor{AScol}{{\bf Accreted stars}}. We describe in detail below how we define each of these accretion types, which we summarise in Fig.~\ref{flowd}.

\subsection{Procedure and classification}

For clarity, we define the following:

\begin{itemize}
\item{} \textbf{Central galaxy:} The main subhalo of each FOF group.
\item{} \textbf{Satellite galaxy:} Any subhalo other than the main subhalo.
\item{} \textbf{Accretion time(s) onto the central galaxy:} The time(s) at which a tracer enters a sphere of radius 30 kpc centred on the potential minimum of the central galaxy (see Fig.~\ref{flowd}); bound subhaloes can accrete onto a galaxy by entering this sphere. A single tracer, can have as many accretion times as it enters this sphere, however the time of {\it first} accretion is used for the classification of accretion sources, described below. 
\end{itemize} 

For each tracer particle, we determine the distance to the central galaxy at each snapshot (defined as the potential minimum), and identify all snapshots that precede the first instance a tracer accretes onto the main galaxy - the first time a tracer is within a 30 kpc radius of the central galaxy (see Fig.~\ref{flowd}). The ISM of galaxies is indicated by the grey shaded regions At each snapshot, we determine: the type of element in which a tracer particle resides; the subhalo to which it is bound (if any); and number of times that a tracer particle has been wind-recycled. This information enables us to assign each tracer particle to one of the following accretion sources:

\begin{itemize}
    \item \textcolor{IGMcol}{{\bf Smooth IGM}}: The tracer has never resided in star-forming ISM gas before accretion, and experiences no wind recycling events before forming a star particle in the central galaxy. We note that this includes both unbound gas and non-star-forming gas belonging to the CGM of subhaloes prior to accretion onto the main galaxy. This choice is reasonable given that the distribution of this gas is typically extended and loosely bound to its subhalo, which may become unbound upon accretion onto the central galaxy.
    \item \textcolor{FFcol}{{\bf Fountain flow}}: The tracer has never resided in star-forming ISM gas {\it before accretion, but subsequently enters at least one wind-recycling event} (see lower-left part of Fig.~\ref{flowd}).
    \item \textcolor{AScol}{{\bf Accreted stars}}: The tracer resides in a star particle born in a subhalo other than the central galaxy.   
    \item \textcolor{SIcol}{{\bf Satellite ISM}}: The tracer accretes onto the central galaxy in the form of ISM bound to a subhalo (upper-left part of Fig.~\ref{flowd}).
    \item \textcolor{SWcol}{{\bf Satellite wind}}: The tracer left the ISM of a subhalo in the form of a wind particle after the snapshot at which it was last part of the subhalo ISM and before accretion onto the central galaxy. 
    \item \textcolor{SScol}{{\bf Stripped satellite}}: The tracer left the ISM of a subhalo \citep[via tidal and/or ram pressure stripping, e.g.][]{SGG17,DNF19} without entering a wind phase (upper-right part of Fig.~\ref{flowd}).
\end{itemize}

We emphasise that smooth IGM accretion is by construction unrecycled material, whereas fountain flow and satellite wind material is by construction recycled material. However, accreted stars, satellite ISM and stripped satellite gas may be comprised of both recycled and unrecycled material.

\section{Results}
\label{sec:results}

\subsection{Origin of today's stars}

\subsubsection{Mass fractions}

\begin{figure*}
\centering
\includegraphics[scale=2.,trim={1.cm 1.cm 1.5cm 0},clip]{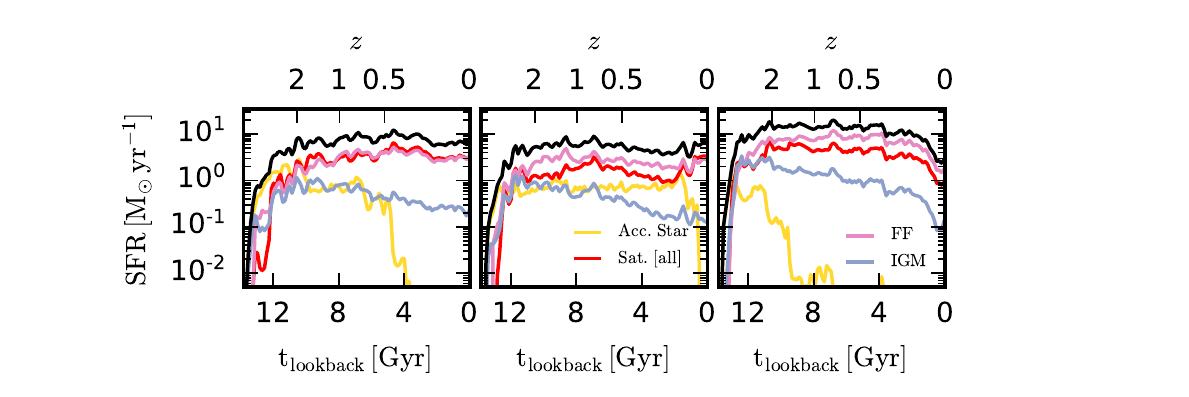}\\
\includegraphics[scale=2.,trim={1.cm 0 1.5cm 1cm},clip]{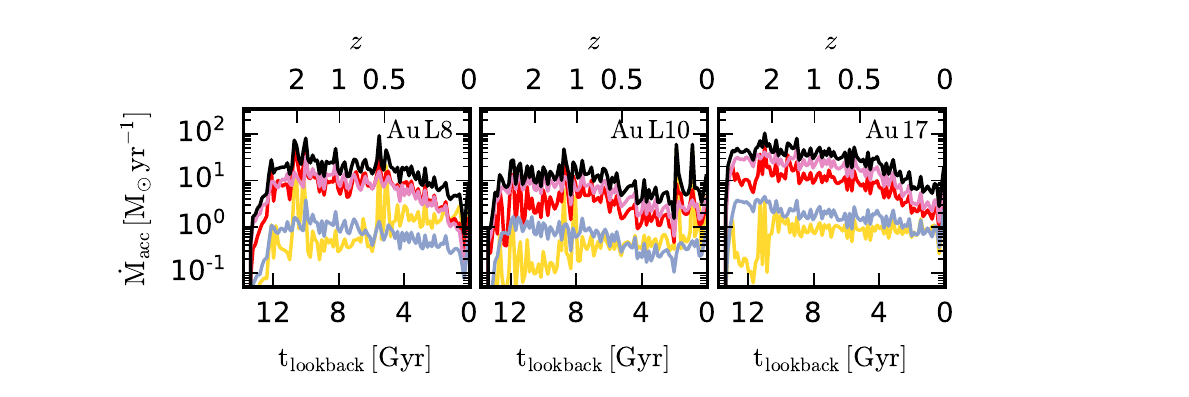}
    \caption{The star formation histories (SFHs, upper panels) and accretion rates (lower panels) of material that ends up in stars at $z=0$ for a selection of simulations. The total SFH and accretion history is denoted by the black curve in each panel, whereas the contributions from different modes of accretion are denoted by coloured curves.}
    \label{fsfh}
\end{figure*}

Fig.~\ref{forig} shows the mass fraction of tracer particles found in star particles at the final time that have been accreted onto the galaxy via each of the modes described in Section~\ref{sec:class} for each simulation. The median mass fraction of each accretion source across the simulation set is summarized in Table~\ref{table1}. In addition, the figure tallies up the total fraction of unrecycled material. One immediately notices that the vast majority of material that ends up in stars has gone through at least one recycling phase; the fraction of recycled material locked in star particles at redshift zero is above $\sim 84\%$ in all simulations. In terms of individual accretion sources, the dominant contribution to wind-recycled material is in the form of galactic fountain flows in most haloes. However, the majority of material associated to satellites (accreted stars, satellite ISM, stripped satellite material and satellite wind accretion) is wind-recycled, and makes up a comparable fraction to that of fountain flows for many haloes. It is important to note, however, that most of the gas originating from satellites does enter what is observed to be a galactic fountain, that is, most of this material behaves in the same way as material that enters a fountain flow after smooth accretion from the IGM (our definition of fountain flow material in the preceding section).

The contribution of satellite wind material ranges from $\sim 15\%$-$30\%$ (with a median of $20\%$), and is always higher than the fraction of material stripped from the ISM of satellites and almost always higher than the fraction of material brought directly into the galaxy in the ISM of satellites. The median fraction of satellite wind $+$ satellite stripped material is $27\%$, and is approximately $10\%$ lower than the values reported in \citet{AAF17} for their 3 Milky Way mass galaxies. This may be because the FIRE simulations analysed in that study employ a more explosive feedback model \citep{HKO14,SHF16} in comparison to Auriga, which would help evacuate gas from satellite galaxies. Another possible explanation for these differences may be the different methods adopted to classify this type of accretion. It may also be the case that small number statistics of \citet{AAF17} show a biased contribution of satellite wind transfer to the final stellar mass budget; 2 of their galaxies show wind transfer fractions within the range of values presented here, therefore a larger sample may be more consistent with our results. 

The median mass fraction of accreted stars is $12\%$ (and can be as little as $1\%$ for individual haloes), which is consistent with values obtained from the Illustris large cosmological box simulation \citep[][]{RPS16}. A notable exception is AuL6, which experiences a major merger at the very recent lookback time of $\sim 2$ Gyr, reflecting the diverse merger histories of the sample. 

\begin{table}
\caption{Table of median accretion fractions for each accretion source (1st column) for tracers that belong to star particles at $z=0$ (2nd column), those that belong to particles on high circularity disc-like orbits (3rd column) and those that belong to counter rotating orbits from the spheroidal component of the galaxy (4th column). Accreted stars (highlighted in bold) show the largest fractional variation between the disc and bulge components.}
\centering
\begin{tabular}{c c c c}
\\\hline
Source & median [all] & median [$\epsilon > 0.7$] & median [$\epsilon < 0$]  
\\\hline
FF & 0.33 & 0.36 & 0.31 \\
Sat. strip & 0.07 & 0.07 & 0.06 \\
Sat. wind & 0.20 & 0.22 & 0.17 \\
Sat. ISM & 0.11 & 0.13 & 0.10 \\
Acc. star & {\bf 0.12} & {\bf 0.02} & {\bf 0.27} \\
IGM & 0.08 & 0.07 & 0.08 
\\\hline
\end{tabular}
\label{table1}
\end{table}

Table~\ref{table1} shows also the median mass fractions of each accretion source for a ``disc-like'' and ``bulge/spheroid-like'' sample of star particles. We select these particles according to an orbital circularity, defined to be $\epsilon = \frac{L_z}{L_{z,\rm max}(E)}$, where $L_z$ is the $z$-component of angular momentum of a star particle and $L_{z,\rm max}(E)$ is the maximum angular momentum allowed for the orbital energy, $E$, of the star particle. Star particles inside a 30 kpc radius with $\epsilon > 0.7$ are on near-circular orbits and characterised as ``disc-like'', whereas those with $\epsilon < 0.$ are characterised to be representative of a spheroid \citep[as evidenced in Fig. 7 of][]{GGM17}. Table~\ref{table1} tells us that all accretion sources contribute similarly for bulge and disc components, with the exception of accreted star particles which make up very little of the disc and a quarter of the bulge, reflecting their  in-situ origin \citep[see][for a thorough analysis]{GMG19}.

\subsubsection{Star formation histories and accretion rates}

The upper panels of Fig.~\ref{fsfh} show the star formation history (SFH) for three of the haloes, as well as the contribution of each of the accretion modes (in this figure, we group satellite ISM, stripping and wind transfer material together for clarity). First of all, we note that although the SFHs vary between the simulations, the total SFHs are generically quite flat: they sustain a fairly constant level of late-time star formation from $z\sim 2$ onward, and tend to peak at intermediate times. Both hydrodynamical simulations and semi-analytic models have shown that the early ejection of gas via strong feedback is required to form prominent stellar discs with flat SFHs \citep[e.g.][]{BSG12b,UNO14} and to reproduce the observed fractions of the star-forming disc galaxy population at low-redshift \citep{HWT15}. Simulations have shown also that late-time wind-recycling appears to be required to match the star formation rate of the main sequence and the galaxy stellar mass function \citep{ODK10,DOB11}. Indeed, our simulations seem to reflect this point that the removal and subsequent late-time re-accretion of wind-recycled gas leads to disc-dominated systems with flat SFHs. 

With respect to the accretion sources, accreted stars and IGM accretion have a tendency to dominate the SFH at very early times ($t_{\rm lookback} > 12$ Gyr), with the former often being the most dominant. This is because the very early assembly of the central galaxy is characterised by the merging of many small building blocks of comparable mass, all but one of which are technically classified as subhaloes that merge and add accreted stars to the main progenitor. Shortly after this time, the other accretion sources become important: fountain flows are established very early. At $z < 2$, the contribution of smoothly accreted IGM gas and accreted stars to the SFH dwindles in favour of fountain flow and satellite gas accretion. The majority of the latter type of material enters into a galactic fountain flow after first accretion, which explains its similarly flat SFH compared to the fountain flow. 

The lower panels of Fig.~\ref{fsfh} show the rate at which material from each source is accreted onto the galaxy. To construct these rates, we have calculated the mass that crosses into a sphere of 30 kpc radius per unit time. Accretion rates in general begin to dwindle at roughly $1 > z > 0.5$, whereas the SFR maintains a rather constant value until $z=0$. This means that the material is {\it gradually} consumed by star formation gigayears after the material is accreted, and that stars at $z=0$ are formed from gas confined to a central zone around the galaxy centre, which we quantify in the next section. Sharp peaks in the accreted star source represent times at which subhaloes accrete onto the central galaxy, which are present also in the satellite gas accretion rates. The low-level accretion between the peaks likely comprises small subhaloes and dispersion dominated stellar halo material that traverses the sphere of 30 kpc radius.

\subsection{Wind recycling statistics}

\begin{figure}
\includegraphics[width=\columnwidth,trim={0.1cm 0.2cm 0.4cm 0.5cm},clip]{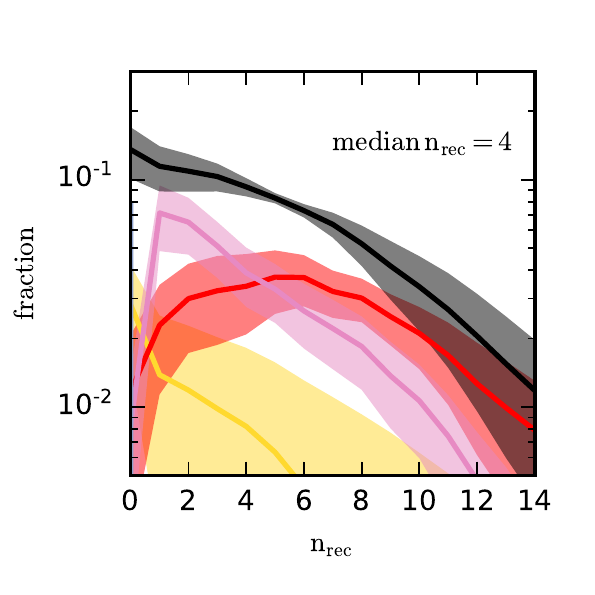}\\
\caption{The fraction of star particles at redshift zero that have formed from gas recycled $n_{\rm rec}$-times, for all simulations: solid curves and shaded regions indicate the median and 1-$\sigma$ scatter averaged across the simulation suite. The total fraction is given by the black curve, whereas the fraction for each accretion mode is denoted by the coloured curves, following the colour scheme of Fig.~\ref{forig} (pink: fountain flow; red: satellite origin; and yellow: accreted stars), but grouping all satellite gas sources together as a red curve. The median number of total recycling times (including all sources) across the suite is 4. Satellite sources have on average undergone more recycling events compared to fountain flows because lower mass galaxies have larger mass-loading factors, therefore satellite material is loaded into winds at a higher rate compared to the central galaxy.}
    \label{frecf}
\end{figure}

\begin{figure}
    \includegraphics[width=\columnwidth,trim={0 1.4cm 0.5cm 1.1cm},clip]{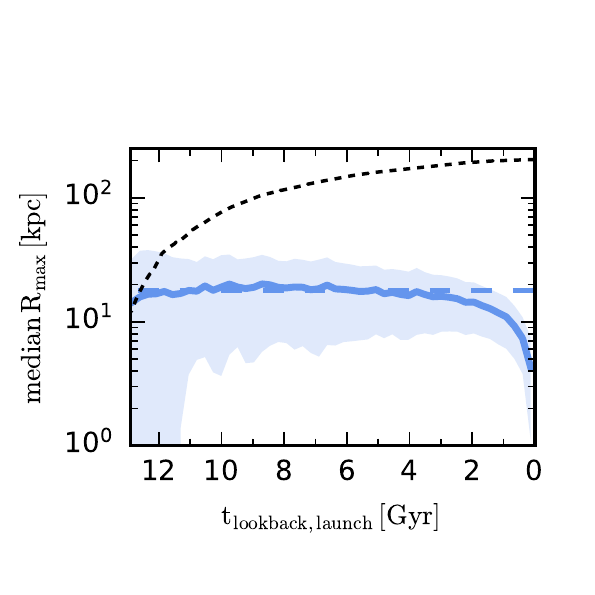}\\
    \includegraphics[width=\columnwidth,trim={0 0.5cm 0.5cm 1.3cm},clip]{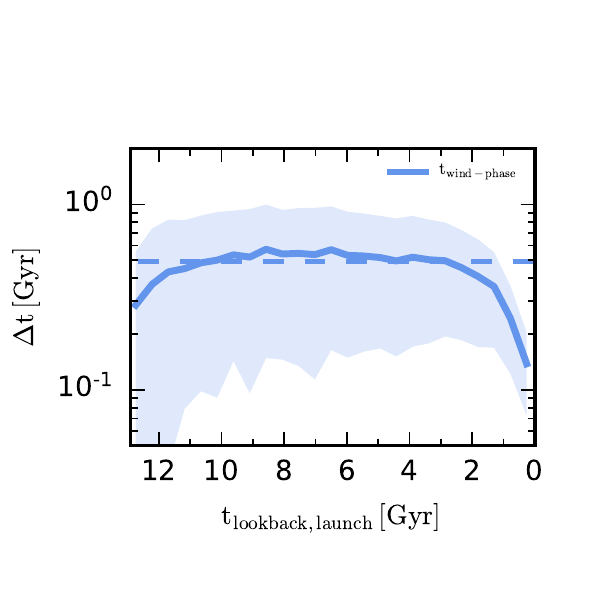}
    \caption{Statistics for the maximum radial distance ($R_{\rm max}$, top panel) attained by fountain flow material after each wind event, and the wind recycling times (bottom panel), $t_{\rm wind-phase}$, defined as the time between subsequent wind phases, for all simulations.  Each quantity is binned according to the time at which a wind is launched. The median values and inter-quartile ranges are indicated by solid curves and shaded regions, respectively. The horizontal dashed lines indicate the median values of $R_{\rm max}$ and recycling times for all wind events of the fountain flow material across the simulation suite. In the top panel, the median virial radius of all simulations is indicated by the black, dashed curve.}
    \label{ffstat}
\end{figure}

Fig.~\ref{frecf} shows the distribution of the number of times the gas elements of each accretion mode have gone through a wind recycling phase. The shape of the distribution is characterised by a quasi-exponential decrease for an increasing number of recycling events. This shape is expected in our model for repeated star formation events: the probability to select cold gas to form a star or wind particle is of exponential form (see equation.~\ref{eqp1}). Satellite sources have on average undergone more recycling events compared to fountain flows because our definition requires fountain flow material to be unrecycled before first infall, whereas this need not be the case for satellite material. Furthermore, lower mass galaxies have larger mass-loading factors, therefore satellite material is loaded into winds at a higher rate compared to the central galaxy. The accreted stars have on average experienced fewer wind recycling events owing to their old age; they have had even less time to be recycled.

In what follows, we focus on the tracers classified as fountain flow only. To understand the timescale on which the fountain flows operate, we calculate $\Delta t_{\rm wind-phase}$, the time between two successive wind-launch times of a tracer. The bottom panel of Fig.~\ref{ffstat} shows the median and spread of $t_{\rm wind-phase}$ for all fountain flow wind recycling events across the simulation suite, as a function of wind launch time, whereas the top panel of Fig.~\ref{ffstat} shows the median and spread of the maximum radial distance fountain flow elements attain. The wind-recycling timescale steadily increases from about 300 Myr to about 500 Myr between $t_{\rm lookback} \sim 13$ Gyr ($z\sim 6$) to $t_{\rm lookback} \sim 8$ Gyr ($z\sim 1$), after which time it stays constant\footnote{The sharp drop in the last 2 Gigayears of evolution reflects the bias of our selection of tracers found in star particles at $z=0$; in order to be locked into a star particle, winds launched at these very late times must necessarily have short $\Delta t_{\rm wind-phase}$ values.}. 

The evolution of these timescales can be understood in terms of our feedback model: for $t_{\rm lookback} \geq 8$ Gyr ($z\sim 1$), the dark matter halo grows (as indicated by the dashed curve in the top panel of Fig.~\ref{ffstat}). This leads to a decrease in mass loading factor\footnote{We use the method of \citet{VQF18} to calculate the mass loading factors: at a given time, we select tracer particles in star-forming gas within 20 kpc of the central galaxy. In another snapshot $\sim 1.5$ Gyr later (approximately the gas consumption time-scale), we divide the total mass of those tracers that are still gaseous by the total mass of those tracers that formed stars.} from $\sim 10$ at $z\gtrsim 1$ to a few at $z\sim 1$, after which time it remains approximately constant. Likewise, the wind velocity increases with decreasing redshift only until $z\sim 2$, and changes little afterwards. While the minimal increase of $R_{\rm max}$ during this early epoch implies that the timescale for expulsion and re-accretion of wind material should be approximately constant, the decreasing mass loading means that the wind-recycled tracer spends more time in the ISM after re-accretion with decreasing redshift (see equations \ref{eqp1} and \ref{eqp2}), thus increasing the time between successive wind-recycling events. This evolution is more prominent at very early times ($t_{\rm lookback} \sim 12$ Gyr) as the halo grows in the inner parts first. After $t_{\rm lookback} \sim 8$ Gyr, the dark halo around the central galaxy stops growing, and the wind model parameters, and thus the $\Delta t_{\rm wind-phase}$, remain constant. We note that with a median of 4 wind-recycling events (as shown in Fig.~\ref{frecf}) and a median $\Delta t_{\rm wind-phase}$ of 500 Myr, the median time between accretion onto the ISM for the first time and star formation is about 2 Gyr, which is consistent with the gas consumption timescale of the Kennicutt star formation timescale \citep{K81}, and well below the age of the universe for z $< 1.5$.  Thus the fountain results in a smoothing of and delay in the SFH, as seen in Fig.~\ref{fsfh}.

\begin{figure*}
    \includegraphics[scale=1.3,trim={0 3cm 0 0},clip]{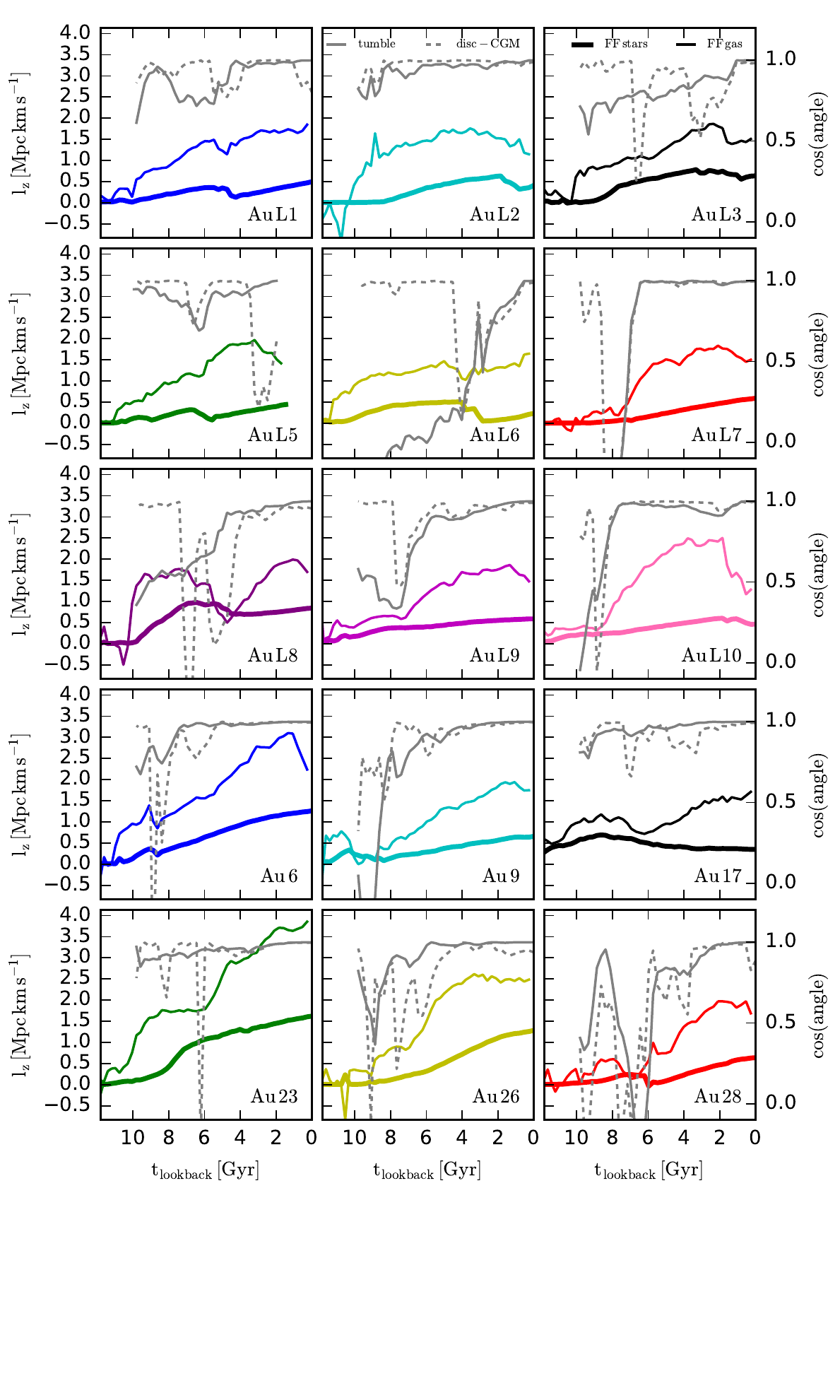}
    \caption{The evolution of median specific angular momentum of tracer particles identified as part of a galactic fountain flow, for each simulation. Tracers particles that are in gas (stars) at a given time are indicated with the thin (thick) lines. The grey solid curves indicate the disc tumbling angle, defined to be the angle between the disc spin axis at a given time and the disc spin axis at $z=0$. The grey dashed curves indicate the angle between the disc spin axis and that of the CGM within a 50 kpc radius, at a given time. The cosine of these angles are marked on the right-hand axis.}
    \label{lzevo}
\end{figure*}

The median recycling timescale found for our galaxies is a factor of 2-3 longer than that found for the Milky Way mass galaxies in \citet{AAF17} and about half of that found in \citet{CDG16} and \citet{TCM19}. We note that we have defined our wind recycling time as the time between two successive wind events, whereas the aforementioned studies define it to be the time between a wind event and re-accretion into the star forming gas. It is therefore expected that our timescales would become more similar to those of \citet{AAF17} were we to adopt their definition, however they would become increasingly disparate with respect to \citet{CDG16}. Although part of the differences may be attributable to different galaxy formation and feedback models, some part of these discrepancies are likely due to uncertainties in the estimated time at which gas re-enters the star-forming regime, which depends on the temporal spacing between simulation snapshot outputs. Indeed, this is why we prefer to define the wind recycling timescale as the time between successive wind events, for which we know the times exactly.

The characteristic zone extending to a few tens of kiloparsecs around the galaxy in which wind-recycling and the galactic fountain operate evolves little with redshift \citep[in rough agreement with Fig. 13 of][]{AAF17}. In our model, the wind velocity grows approximately with the square-root of the gravitational potential, and therefore approximately linearly with the galactic escape speed. This simple argument is in line with the expectation that $\rm R_{max}$ should not evolve much even before the dark halo stops growing at around $z\sim 1$. However, the evolution likely depends non-trivialy on the complexity of the evolving accretion modes and hydrodynamics in the halo gas, obfuscating a truly clear interpretation.

\subsection{Fountain flows and angular momentum}

\begin{figure*}
\includegraphics[scale=1.55,trim={0.5cm 0 0 0},clip]{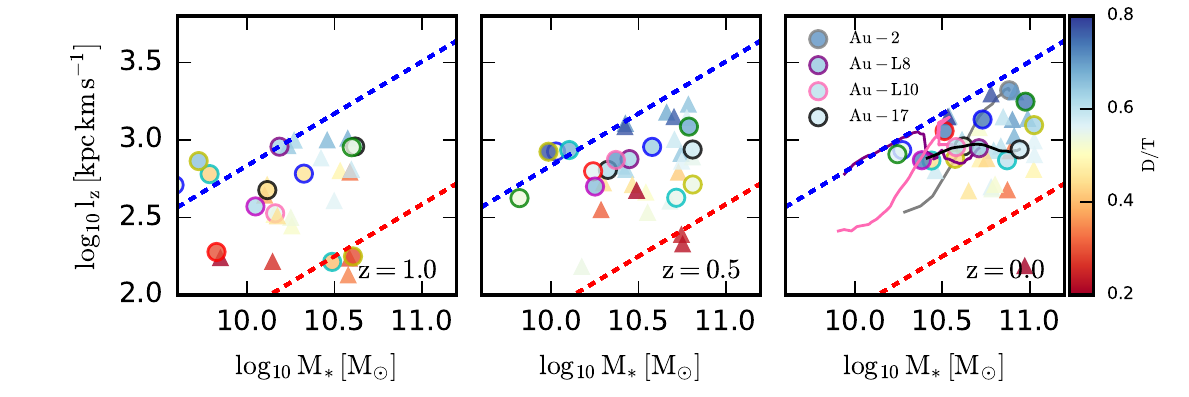}
\caption{Specific angular momentum of stars as a function of stellar mass for each simulated galaxy at $z=1$ (left), $z=0.5$ (middle) and $z=0$ (right). Circles indicate the galaxies presented in this paper, whereas triangles indicate other Auriga galaxies from the original set. The Fall relation \citep{RF12} for disc-dominated and bulge-dominated systems are shown by the blue and red curves, respectively. Evolutionary tracks for three systems for $0 < z < 2$ are drawn on the right-most panel to highlight the effects of: prograde minor mergers (Au 2, Au L10); counter-rotating minor mergers (Au L8); and a secular evolution driven by a strong bar (Au 17).}
\label{fall}
\end{figure*}

In the previous section, we established that the vast majority of star particles at $z=0$ are made from wind-recycled gas, most of which has taken part in a galactic fountain. In this section, we focus on how they affect the evolution of the angular momentum distribution of the simulated galaxies. 

Fountain flows are expected to play an important role in increasing the vertical component of the specific angular momentum, $l_z$, of discs \citep{F17}. To determine whether this holds true, we show in Fig.~\ref{lzevo} the evolution of specific angular momentum, $l_z$, of fountain flow tracer particles in the frame of the disc spin-axis\footnote{The disc spin axis is taken to be the eigenvector of the moment of inertia tensor of star particles within a tenth of the virial radius that most closely aligns with the principal angular momentum axis of the same star particles.}. We divide the fountain flow tracers into two groups: one for tracer particles that reside in star particles (thick curves); and one for tracer particles that reside in gas cells (thin curves), at a given time. Thus, the sum of the two curves corresponds to a fixed set of tracers (those identified as fountain flow in Fig.~\ref{forig}), but shifts from almost all gas on the left to almost all stars on the right, as star formation proceeds to turn gas into stars. 

In most cases, fountain flow gas (thin curves) increases its median specific angular momentum with time, which is converted into increasing stellar specific angular momentum (thick curves) as gas is gradually turned into stars. We find that strong increases in fountain flow gas $l_z$ correlate with good alignment between the disc and surrounding gas. We find also that gas-rich minor mergers can significantly affect this alignment in two distinct manners. Firstly, prograde mergers that quiescently in-spiral \citep[identified as a mechanism to create large discs in][]{GGM17} can drive up $l_z$ of fountain flow gas by: i) torquing the central disc spin axis into alignment with that of the surrounding gas; and ii) providing high angular momentum gas to the CGM from which the fountain can extract $l_z$. Evidence for the former is provided by Au L7 and Au L10 at $t_{\rm lookback}\sim 8$ Gyr, whereas the latter is seen in Au 23: the growth of $l_z$ of fountain flow gas stalls at $t_{\rm lookback}\sim 8$ Gyr owing to a lack of infalling high-$l_z$ gas, until it is provided by a minor merger\footnote{Incidentally, this mechanism is responsible also for the formation of a chemical thin/thick disc dichotomy, similar to that seen in the Milky Way \citep{GBG18}.} at $t_{\rm lookback}\sim 6$ Gyr. This indicates that the fountain must be able to directly mix with high-$l_z$ CGM in order to increase its angular momentum, which we demonstrate below.

Secondly, we find evidence that {\it retrograde} minor mergers act to decrease fountain flow gas $l_z$ in the frame of the disc. As can be seen for halo Au L8 in Fig.~\ref{lzevo}, disc tumbling, disc-CGM misalignment and a decrease in fountain flow gas $l_z$ all coincide at $t_{\rm lookback}\sim 7$ Gyr: the time at which a counter-rotating satellite approaches the central galaxy. The stellar $l_z$ decreases after some lag time as the fountain flow gas is converted into stars. Afterwards, the fountain flow gas $l_z$ increases as wind-recycling establishes a galactic fountain aligned with the CGM. Contrasting the evolution of Au L8 and Au 23, which experience retrograde and prograde minor mergers, respectively, at around the same time, clearly illustrates the dependence of the orbital orientation of minor mergers on the final $l_z$ of gas and stars. That prograde (retrograde) minor mergers increase (decrease) the angular momentum of the gas around galaxies is generally consistent with results from the EAGLE project \citep{LSB18}.

Aside from minor mergers, other important factors for the stellar $l_z$ evolution include: violent major mergers and secular evolution. Whereas the former cause direct losses to the specific angular momentum of stars and gas (Au L6, Au 28), the latter cause gradual losses to the stellar $l_z$ in particular, which is the case for the isolated halo Au 17. This halo develops a strong bar at $z\sim 1$ \citep[see Fig. 5 of][]{GSG16}, which drives secular changes in the shape of the $l_z$ distribution such that the median of the distribution decreases. The effects are strong: even though the galactic fountain (thin curve) contains more angular momentum than the stellar disc (thick curve), it cannot override the scattering effects induced by a strong bar by replenishing the stellar distribution with high-$l_z$ stars. This highlights the high degree of complexity of physics involved in disc growth that can be captured by our cosmological simulations (see Fragkoudi et al. in prep for the properties of Auriga barred galaxies.).

\begin{figure*}
    \includegraphics[scale=0.25,trim={0 0 4.8cm 0},clip]{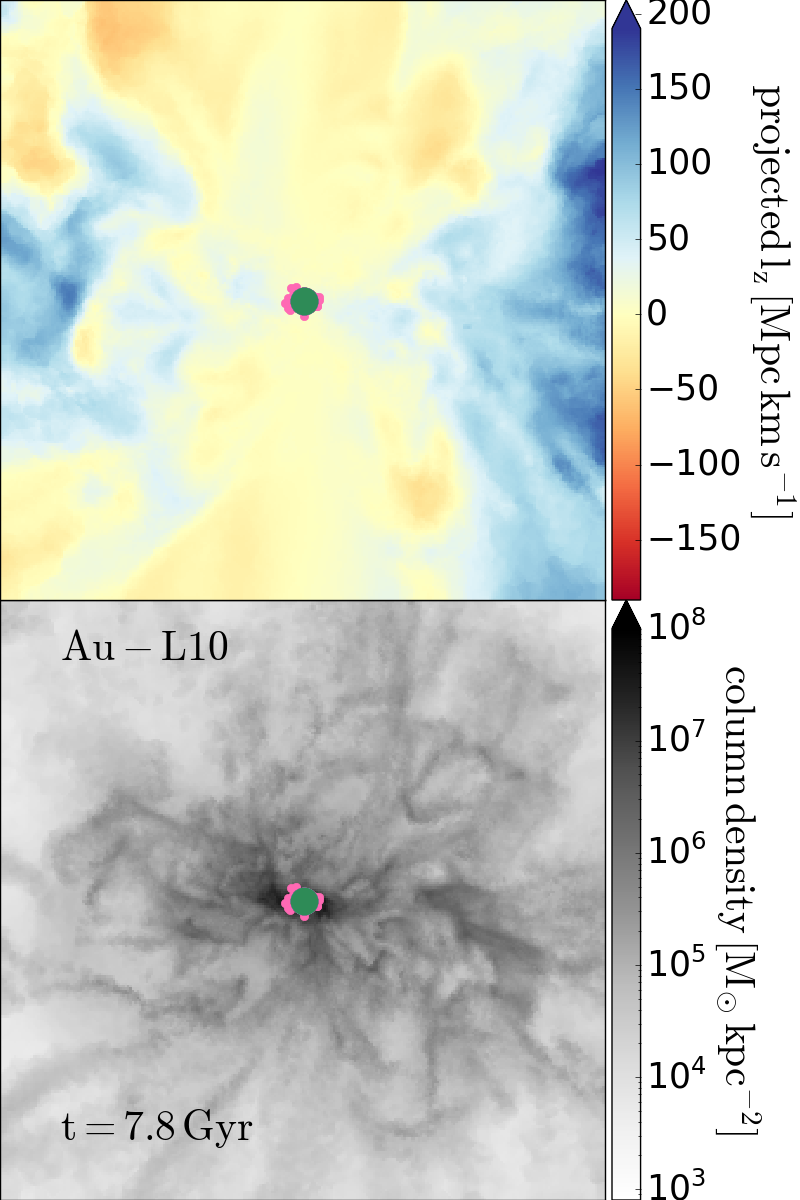}
    \includegraphics[scale=0.25,trim={0 0 4.8cm 0},clip]{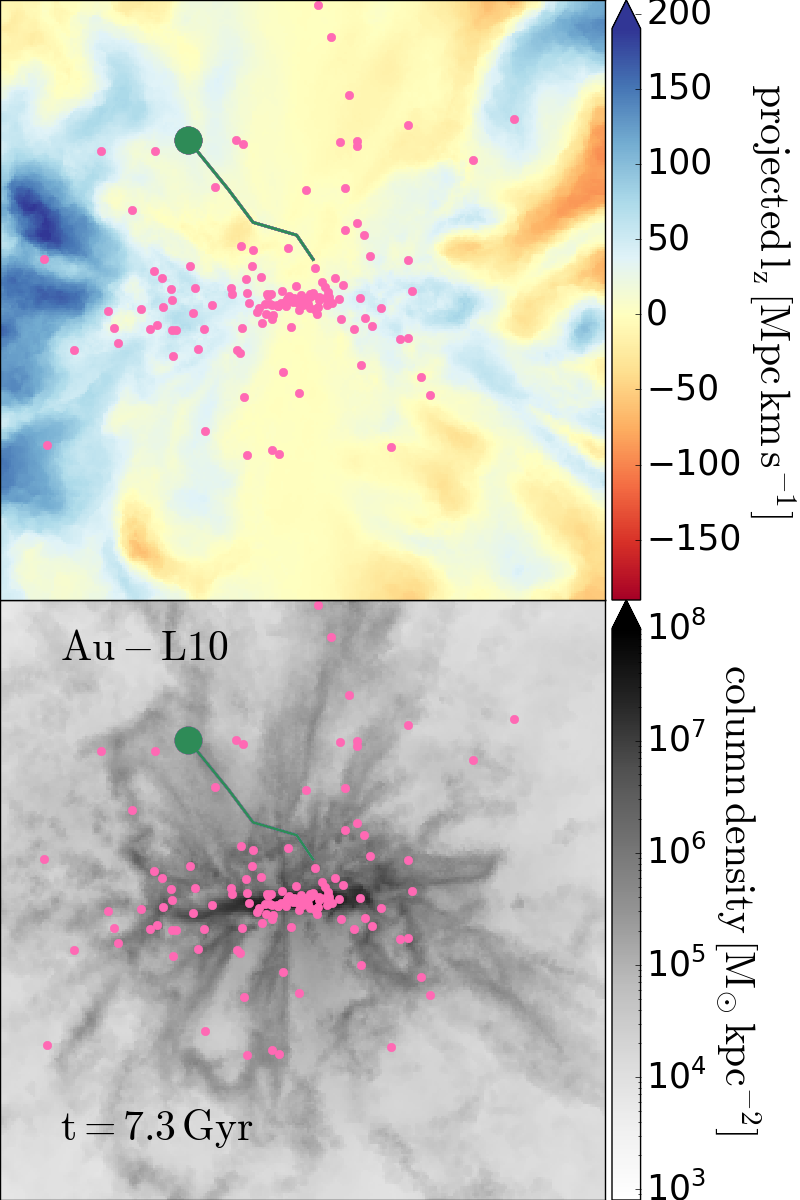} 
    \includegraphics[scale=0.25,trim={0 0 4.8cm 0},clip]{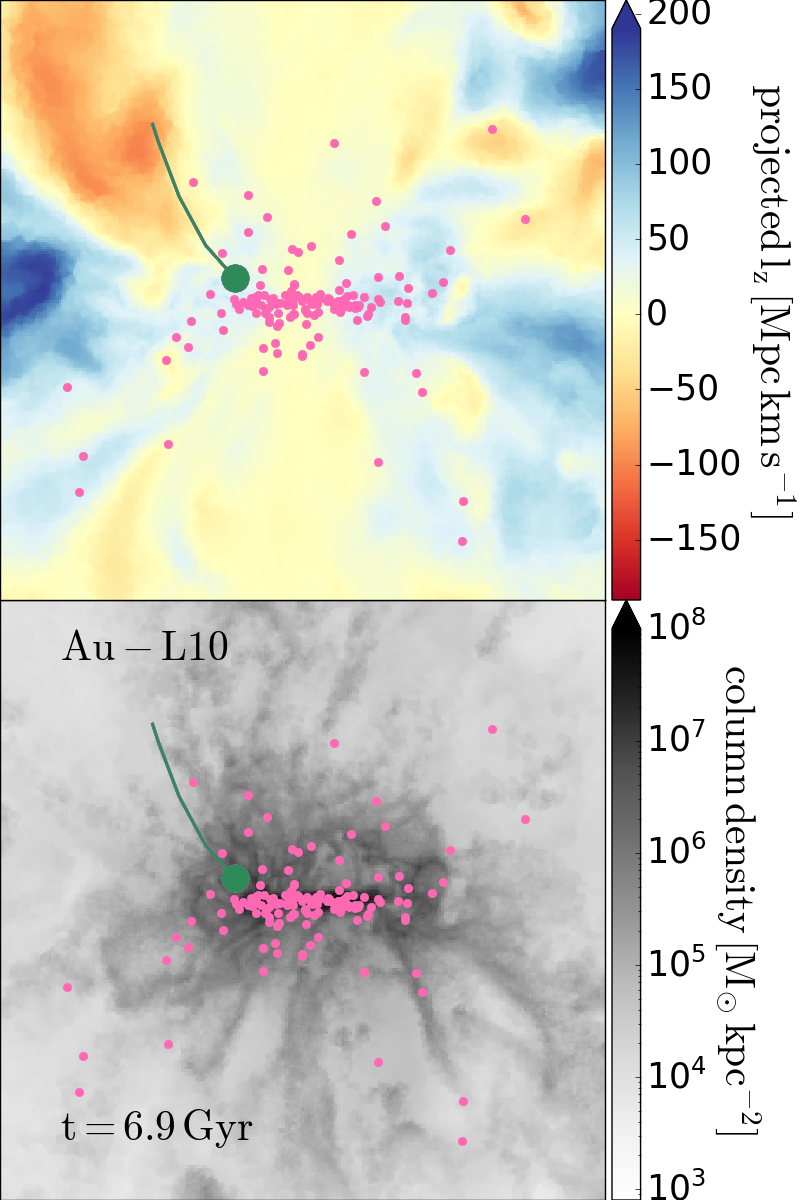} 
    \includegraphics[scale=0.25,trim={0 0 0 0},clip]{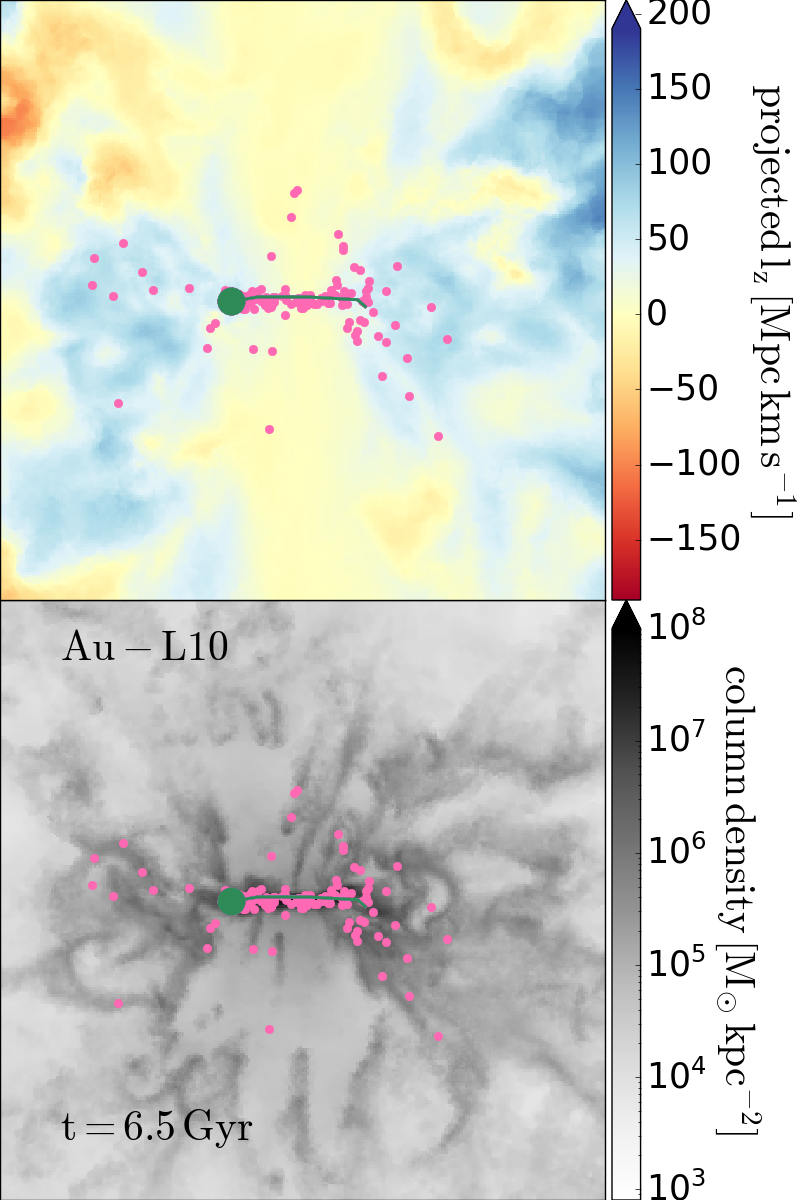} 
    \includegraphics[scale=0.34,trim={0 1.5cm 0 0},clip]{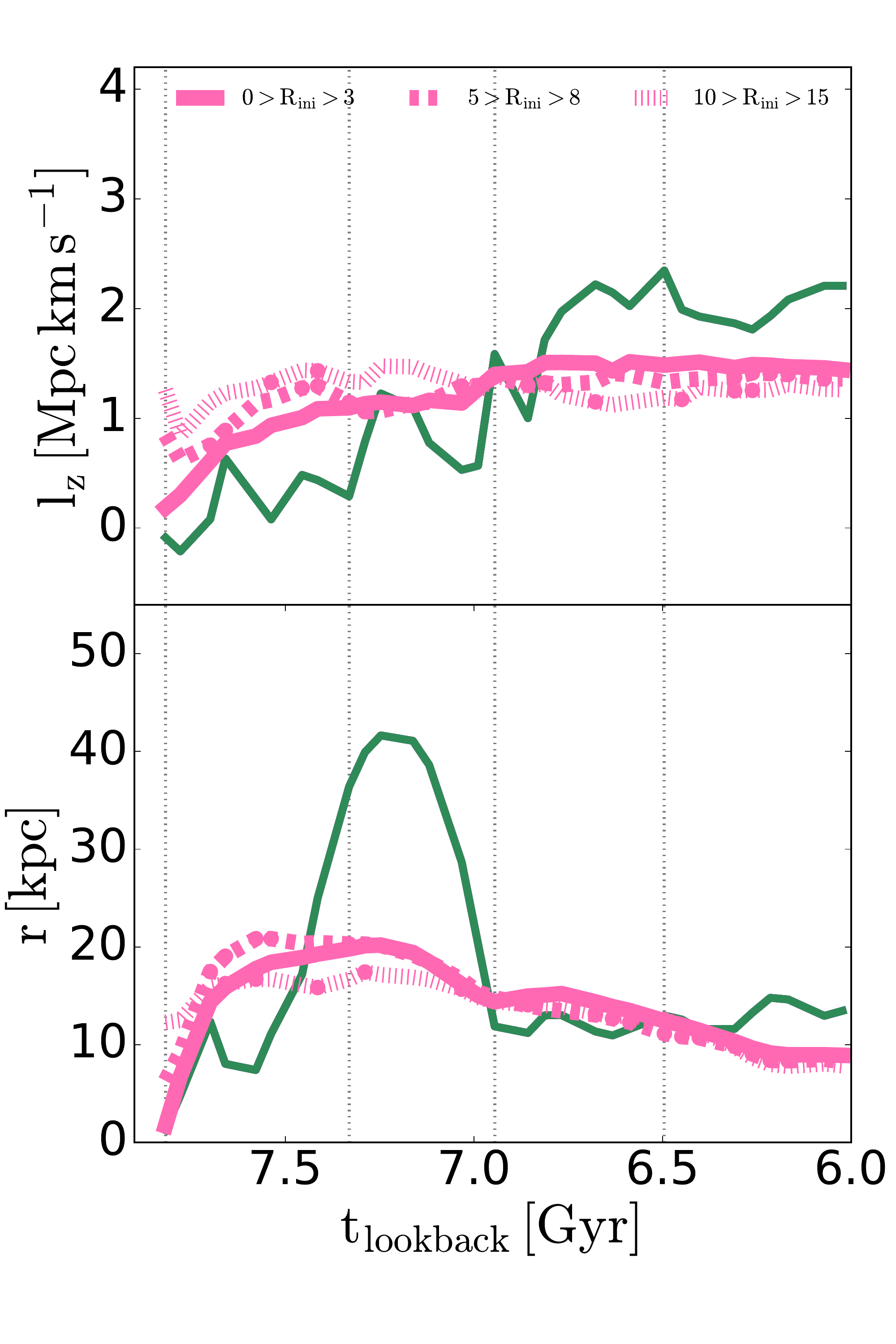}
    \includegraphics[scale=1.35,trim={0 0.8cm 0.5cm 0},clip]{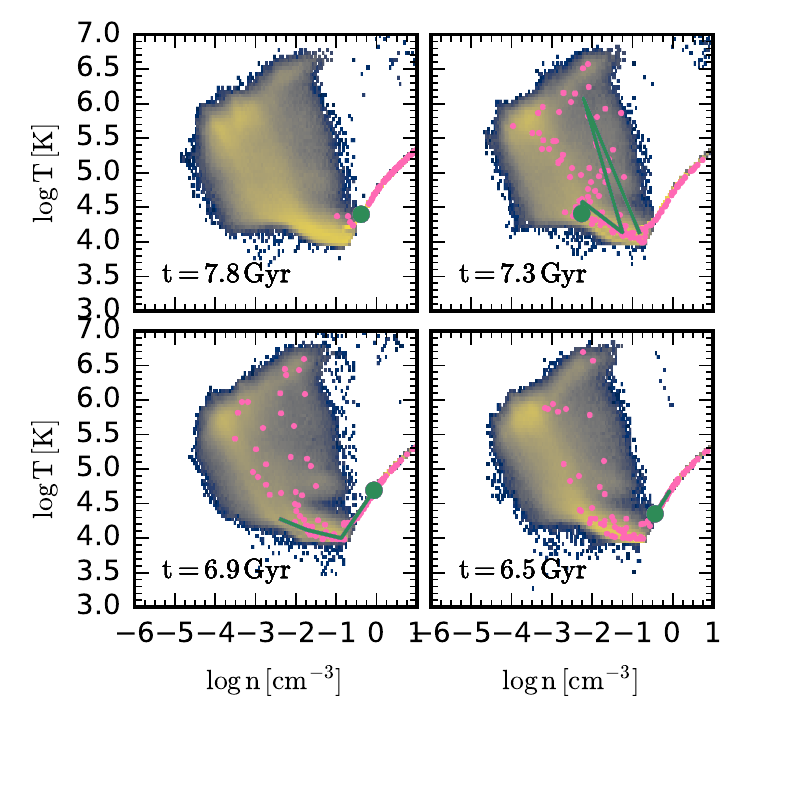}
    \caption{{\it Upper panels} (a-d): Sequence of 10 kpc-slice projections ($100\times100$ kpc) of gas specific angular momentum (top) and gas column density (bottom) for simulation Au L10. Also shown is the evolution of a sample of fountain flow tracers (pink dots) selected to be within 3 kpc at $t=7.8$ Gyr (panel a) and to have entered a wind-recycling phase within the next 60 Myr. The panels highlight the position and 300 Myr past trajectories of an example particle to highlight how angular momentum is acquired from the CGM (green symbols/lines). {\it Lower-left panels}: Evolution of specific angular momentum (top) and spherical radius (bottom) of the fountain flow particles shown above (solid pink curve), in addition to samples of particles that have radii $\rm 8 > R_{ini} > 5$ kpc (dashed curve) and $\rm 15 > R_{ini} > 10$ kpc (dotted curve) at $t=7.8$ Gyr, and the example particle (green curve). Times corresponding to the projections in the upper panels are indicated with vertical dotted lines. {\it Lower-right panels}: Phase diagrams for all gas within 100 kpc of the galactic centre (shaded 2D histogram) at each of the times shown in the upper panels, showing the sample of fountain flow tracers shown in the panels above and the example particle. Fountain flow material is first expelled from the multi-phase, star-forming ISM, and mixes with relatively low-density, high temperature CGM gas ($\rm log(n) \sim -2$, $\rm log(T) \sim 6$-$6.5$) with high specific angular momentum. The gas cools rapidly, undergoes temperature fluctuations around $\rm log(T)\sim 4$-$5$ as it mixes with gas during its re-accretion, before settling into the ISM with higher angular momentum and galacticentric radius.}
    \label{ffmech}
\end{figure*}

\subsection{The angular momentum - stellar mass relation}

The relationship between the specific angular momentum and stellar mass of a galaxy \citep[the Fall relation,][]{RF12} has been used as a quantitative definition of galaxy morphology, to classify bulge- and disc-dominated galaxies, respectively. \citet{FR18} find that galaxies with high disc-to-total mass ratios (D/T) follow a linear relation for which the $l_z$ of the stellar disc and dark halo are approximately equal (the disc relation), whereas galaxies with low D/T (or bulge/spheroid systems) lie on a parallel relation with lower $l_z$ (the bulge relation).  

To determine whether our simulated galaxies follow these trends, we calculate the stellar mass and $l_z$ using all star particles within a ``galaxy radius'' (defined to be one tenth of the virial radius at each time, which accommodates the central galaxy). For the D/T ratios, the total mass, T, is taken to be the stellar mass within a galaxy radius. The disc mass, D, is taken to be a linear average of the two kinematic definitions described in \citet{GGM17}, which bracket the true disc mass: i) the mass of star particles with $\epsilon > 0.7$; ii) the remaining stellar mass after subtraction of twice the mass of counter-rotating stars from the total mass. We stress that this definition of D/T is intended to provide a rough estimate only of the ratio of disc mass to total mass to assess the relative trends in the Fall relation, and not to reproduce the precise definitions used in observations, which are measured differently. 

To increase statistics, we show in Fig.~\ref{fall} all 40 of the Auriga galaxies \footnote{We do not require tracer particles for this calculation, therefore we extend the sample of simulations to include the complete Auriga simulation suite.} at redshift 1, 0.5 and 0 on the Fall relation, coloured by D/T. In each panel of Fig.~\ref{fall}, we show the $l_z$-$M_*$ relations for discs (blue line) and bulges (red line) derived in \citet{FR18}. As in that work, we find that systems with larger D/T lie closer to the disc relation, and systems with lower D/T lie closer to the bulge relation. We emphasize that both the trend and scatter of the simulated galaxies is similar at all redshifts considered, supporting claims that the Fall relation evolves little with redshift \citep{MFP19}.  

To understand how individual galaxies evolve in this plane and shape the overall relation, we highlight the evolutionary tracks of Au 2, Au L8, Au L10, and Au 17, which exhibit distinct evolutions. Firstly, the tracks clearly show that quiescently evolving galaxies in which fountain flows extract angular momentum from the CGM evolve parallel to the disc relation (Au L8 for $1<z<2$) or on even steeper tracks (Au 2, Au L10) than the disc relation if prograde minor mergers are involved. On the contrary, Au L8 evolves off the relation during the period $0.2<z<1$, which coincides with a retrograde minor merger that leads to subsequent disc tumbling and decreases in $l_z$ of the galactic fountain in the frame of the disc. The last $\sim 2$ Gyr of evolution for Au L8 shows evidence of a resumption of disc growth, which corresponds to the time at which the disc and galactic fountain become realigned. Secondly, the track of Au L10 at very late times shows a sharp decrease in $l_z$, owing to a violent merger. The effect of mergers lowering the $l_z$ of galaxies has been noted in previous work \citep[e.g.][]{LTS16,SCF17}, and occurs in many of our simulated galaxies. Thirdly, the strongly barred simulation of Au 17 slowly evolves away from the disc relation; the specific angular momentum decreases slightly as stellar mass increases.  

Thus, we conclude that fountain flows, mergers and internal secular evolution all play a role in shaping the Fall relation over cosmic time, driving individual galaxies up and down the relation to maintain a roughly constant trend and scatter.

\subsection{The mechanism of angular momentum acquisition}

\begin{figure}
\includegraphics[scale=0.5,trim={0.5cm 2.cm 2.5cm 1.6cm},clip]{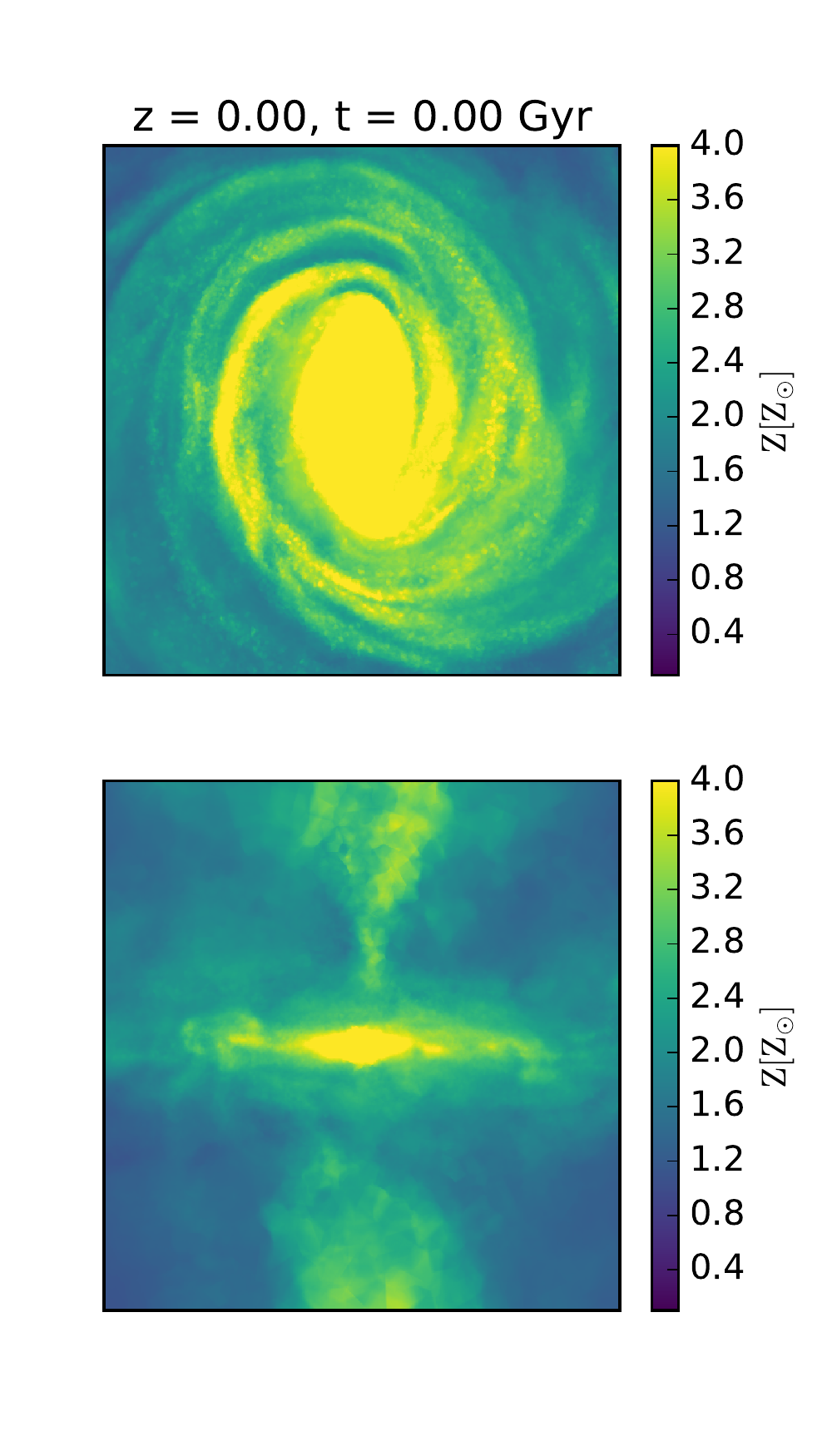}
\includegraphics[scale=0.5,trim={0.5cm 2.cm 0 1.6cm},clip]{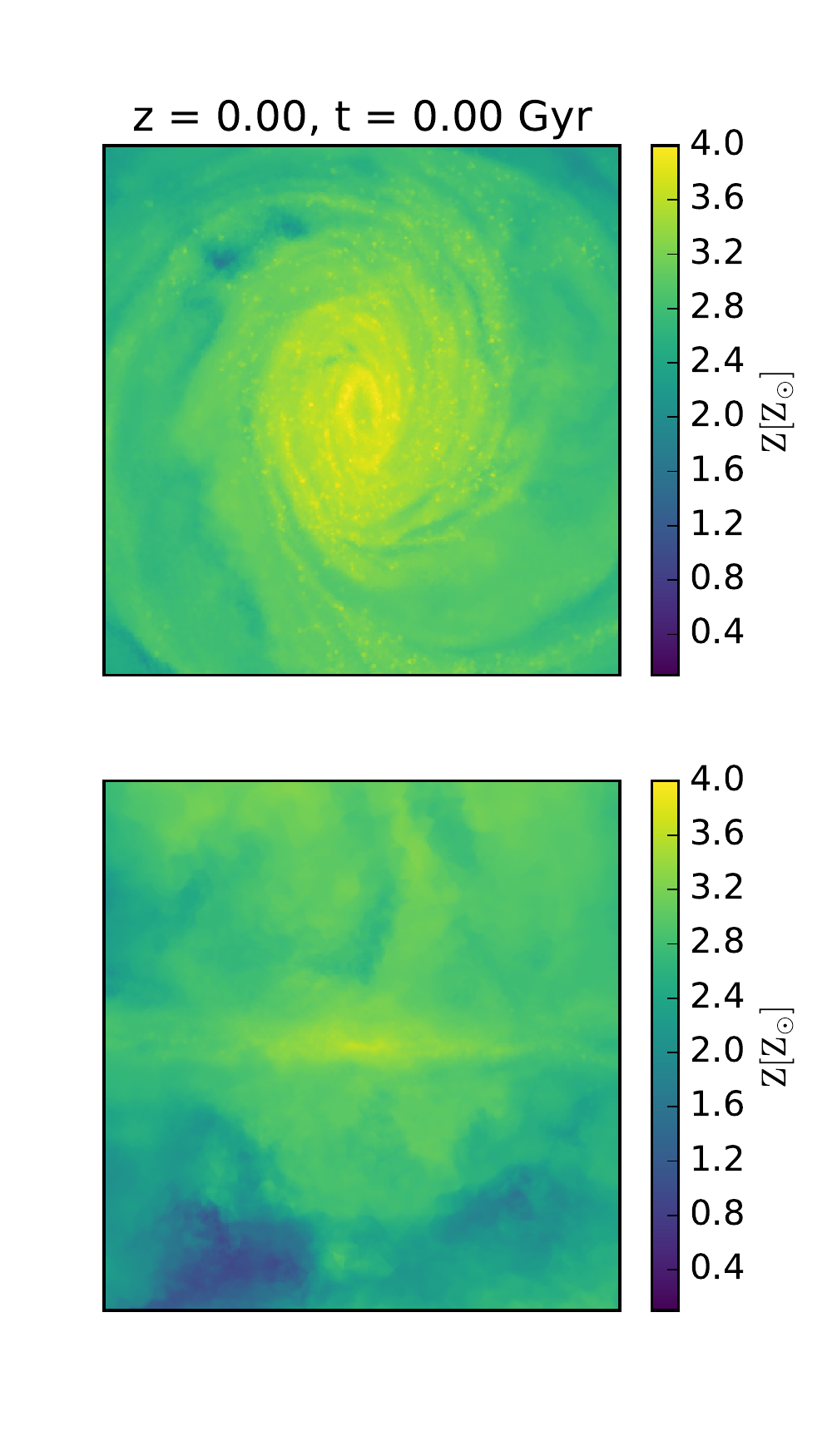}
\caption{Face-on and edge-on metallicity-weighted gas density projections for the fiducial Au 6 with 40\% of metals loaded into winds (left panels) and a re-run with 100\% of metals loaded into winds (right panels). Metals are far more homogeneous in the right panels than the left panels.}
\label{gmetproj}
\end{figure}

\begin{figure*}
\includegraphics[scale=1.3,trim={1.5cm 1.1cm 0 0},clip]{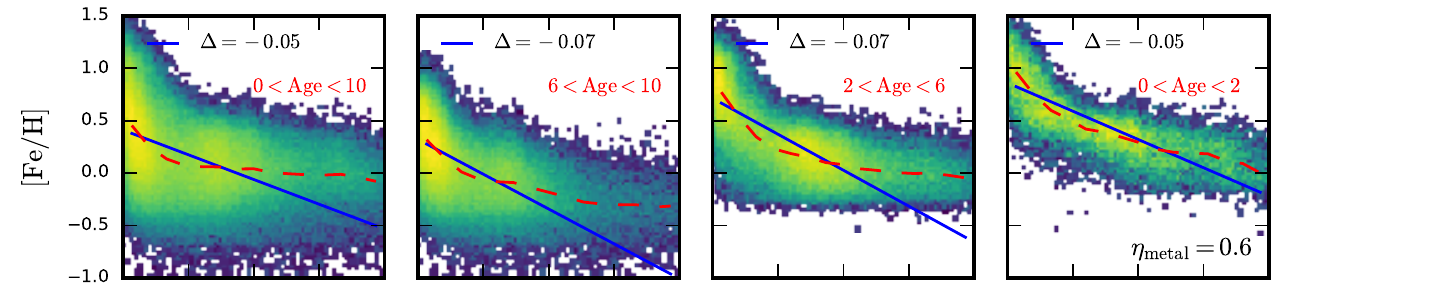}
\includegraphics[scale=1.3,trim={1.5cm 0.3cm 0 0},clip]{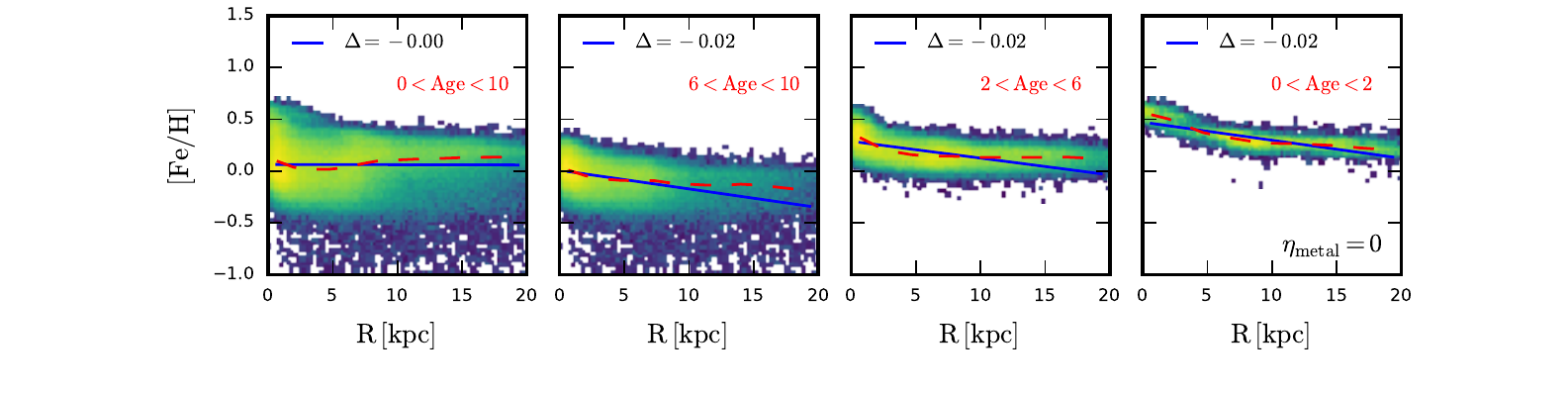}
\caption{Radial metallicity distribution of all stars within 5 kpc of the disc midplane (left panels); stars between 6 and 10 Gyr old (second column); stars between 2 and 6 Gyr old (third column) and stars younger than 2 Gyr (right column) for the fiducial Au 6 (top row) and the re-run with 100\% metal loaded winds (bottom row). The mean metallicity as a function of radius is given by the red dashed lines. Linear fits to the mean metallicity as a function of radius is indicated by the blue solid lines, and the gradient is indicated in the top-left corner of each panel. Radial metallicity gradients (dispersions) are flatter (narrower) for the latter case compared to the fiducial case, owing to the removal of metal-rich gas from the central galaxy to the outskirts and increased metal mixing in the CGM relative to the disc.}
\label{radmet}
\end{figure*}

In this section, we investigate the source and physical mechanism by which fountain flows acquire angular momentum. For clarity, we focus on the relatively isolated halo Au L10, and select a sample of fountain flow tracers at $t_{\rm lookback}=7.8$ Gyr that are within a galactocentric radius of 3 kpc, with low specific angular momentum, and that are about to enter a wind-recycling phase. The spatial evolution of this sample of tracer particles is shown at a sequence of snapshots in the upper panels of Fig.~\ref{ffmech}, which are projections of the gas specific angular momentum (upper row) and gas column density (lower row). The lower-left panels of Fig.~\ref{ffmech} depict the evolution of the three-dimensional radius and specific angular momentum of the particles. The lower-right panels show phase diagrams of gas within 100 kpc of the galactic centre, with fountain flow particles over-plotted. 

To help understand how fountain flows acquire angular momentum from the CGM, we highlight an example particle and its past 300 Myr evolutionary trajectory (green symbols and curves). Shortly after $t_{\rm lookback}=7.8$ Gyr, the particles are ejected from the multi-phase, star-forming ISM\footnote{We note that although our model launches winds isotropically, winds predominantly take the path of least resistance and produce conical bi-polar outflows, as can be seen in the right-most panels of the projected gas density in Fig.~\ref{ffmech} and discussed in \citet{NPS19}.}, and mix with relatively low-density, high temperature CGM gas ($\rm log(n) \sim -2$, $\rm log(T) \sim 6$-$6.5$). This gas has higher $l_z$ than the fountain flow particles (blue regions in the upper panels). As the metal-enriched ejected material mixes with and entrains CGM gas, the angular momentum exchange results in an increase (decrease) in $l_z$ of the ejected material (CGM gas). The increased metallicity encourages the mixed gas to cool \citep{MBF10,MFN11,FMM13}, effectively distilling high-$l_z$ gas from the CGM, which accretes onto the disc. As a result, the tracer particle sample on average increases its specific angular momentum (quantified in the lower-left panels of Fig.~\ref{ffmech}) and settles into the star-forming ISM in the outer-disc: this is clearly seen in panel d of Fig.~\ref{ffmech}, which reveals the orbits of the sample to be extended and near the disc plane. This analysis supports the idea that fountain flow material acquires angular momentum via mixing with the CGM.

The lower-left panels of Fig.~\ref{ffmech} show also the evolution of tracers ejected from intermediate radii ($\rm 8 > R_{ini} > 5$ kpc) and the outer disc ($\rm 15> R_{ini} > 10$ kpc). All three groups of tracers end up with the same specific angular momentum at $t_{\rm lookback}=6$ Gyr: winds launched from the disc at larger radii have larger angular momentum prior to ejection but gain less compared to those ejected at smaller radii. In particular, those ejected from the outer disc exhibit little change. Inside-out formation of our simulated galaxies means that the average radius of wind-ejection sites increases with decreasing redshift, which may contribute to a gradual reduction in angular momentum growth of the galactic fountain at late times seen in some galaxies in Fig.~\ref{lzevo}, e.g., Au L10.

It is interesting to note from the lower-right panels in Fig.~\ref{ffmech} that fountain flow material may fluctuate in density, temperature and $l_z$ during its excursion into the CGM (illustrated clearly by the example tracer in the figure, particularly at $t_{\rm lookback} = 7.3$ Gyr). Indeed, the CGM is a highly structured medium, characterised by filamants and patches of cold and hot gas of high, low and even counter-rotating angular momentum, which seems to evolve on 1 Gyr timescales.

\subsection{Fountain flows and metal mixing}

The last section demonstrated that galactic fountain material increases its angular momentum through mixing with the CGM. Given that the galactic fountain is powered by wind-recycled, and therefore metal-enriched, material, it follows that fountain flows play an important part in setting the metal distribution of the galaxy. In this section, we briefly highlight the qualitative effects of galactic fountains on the metal content of the CGM, and the radial metallicity distribution of disc stars. 

For a clear illustration of the effects of galactic fountains on the metal distribution, we re-run halo Au 6 with a metal loading factor $\eta _{\rm metal} =  0$. This ensures $100 \%$ of metals are carried away by winds (see Section~\ref{sec:sim}), compared to the fiducial value of $\eta _{\rm metal} = 0.6$, for which $40\%$ of the metals are carried away by winds. In Fig.~\ref{gmetproj}, we show face-on and edge-on projections of the gas mass-weighted metallicity for the fiducial case and the $\eta _{\rm metal} =  0$ case. The difference between the two simulations is dramatic: the fiducial case (left panel) shows a clearly metal rich disc, with clear azimuthal inhomogeneities that appear to trace a barred-spiral structure. The CGM, on the other hand, is relatively metal poor, except for metal enriched outflows that trace a conical structure (see Pakmor et al in prep). For the simulation with fully metal loaded winds (right panel), the disc is relatively metal poor in comparison, and more similar to the CGM. These striking differences indicate that fountain flows are crucial for setting the metallicity distribution of gas in the galaxy.

The strong effect of fountain flows on the metal distribution of gas in the galaxy illustrated above manifests also in the stellar metallicity distribution. Fig.~\ref{radmet} shows the radial metallicity distribution of all stars (left panels) and grouped by age (old stars, intermediate stars and young stars are shown in the second, third and fourth panels, respectively). There are two important differences that are most apparent in the youngest stellar group: i) the radial metallicity gradient is shallower and ii) the dispersion at a given radius is smaller for the $\eta _{\rm metal} =  0$ case (lower panels) compared to the fiducial case (upper panels). The flatter gradients reflect the effectiveness of the fountain flow to convert low angular momentum, metal rich material into high angular momentum material, which both decreases the central metallicity and increases the outer disc metallicity relative to the fiducial case \citep{TCK12,GPB13,ADO14,MHF17}. The second difference, i.e. the metallicity dispersion at a given radius, indicates that metal mixing is more efficient in the CGM than in the disc; fountain flows pollute the CGM with metals that mix efficiently before falling onto the outer disc. We note that these trends are observed for the older stellar populations (second and third panels of Fig.~\ref{radmet}), however the metallicity dispersion is larger owing to dynamical effects such as mergers \citep{RFG16,BSS18} and radial migration that broaden the metallicity distribution (\mbox{\citealt{MCM13}}, \citealt{DiM13}, \citealt{GKC15}, \citealt{GSK16}, \citealt{SSK16}, \citealt{LDN16}, \citealt{KGG17}). We defer a detailed analysis of the drivers of the stellar metallicity distribution to future work (Tronrud et al in prep).

\section{Convergence: number of tracers and resolution}
\label{sec:conv}

\begin{figure*}
\centering
\includegraphics[scale=2.,trim={1.cm 1.cm 2.2cm 0},clip]{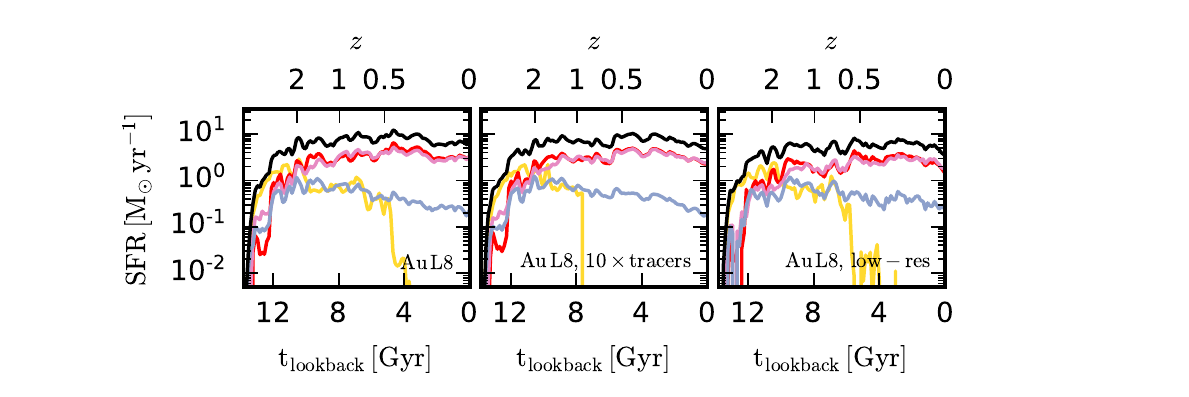}\\
\includegraphics[scale=2.,trim={1.cms 0 2.2cm 1.cm},clip]{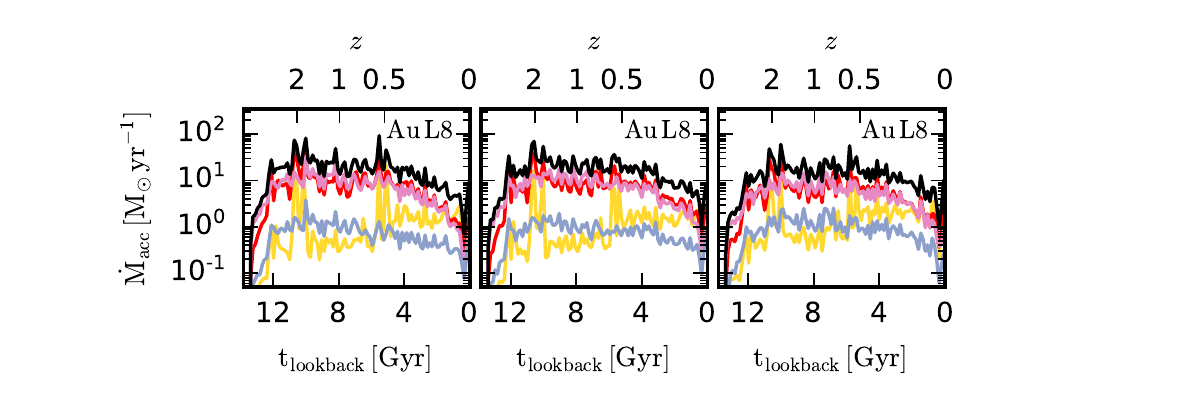}
\caption{As Fig.~\ref{fsfh}, for the fiducial halo Au L8 (left panel), the same halo simulated with 10 times as many tracer particles (middle panel) and the same halo simulated at a factor 8 poorer mass resolution with 1 tracer particle per baryonic element (right panel).}
\label{app2}
\end{figure*}

Owing to the Monte-Carlo approach of sampling the outward facing fluxes across cell faces, the fluid flow is expected to be traced more faithfully for a larger number of tracer particles employed \citep{GVN13}. Here, we verify that our results do not depend on the number of tracer particles by re-running simulation Au L8 with 10 times more tracer particles per gas cell than the fiducial Au L8 presented in this paper. In Table~\ref{tab} and Fig.~\ref{app2}, we compare the categorisation, star formation histories and accretion rates of these haloes (analagous to Figs.~\ref{forig} and ~\ref{fsfh}). Clearly, these plots are reproduced at the quantitative level. Note that the SFH and accretion rate of the run with 10 times as many tracers per gas cell appear slightly smoother in comparison to the fiducial run, in line with the expected reduction of Poisson noise of a factor of $\sim 3$. We therefore conclude that our results are robust to the number of tracer particles employed.

One concern relating to the robustness of the results presented here will be the numerical resolution around the disc-halo interface: the adaptive resolution of our simulations means that the spatial resolution of gas cells in the CGM is of order several kiloparsecs. The implications of this relatively coarse spatial resolution in the CGM for galaxy evolution are not yet well known, but it is expected that higher resolution is required for accurate treatment of hydrodynamical flows and mixing, ram-pressure stripping of satellites, feedback and the formation of cold clouds \citep[e.g.][]{VSM19,SPV19,HSH19,PCT19}. Of particular concern is the scale-independent (above the thermal conduction scale) nature of the thermal instability, which may cause the gas to become ever-more clumpy with increasing resolution with different mixing and transport properties in comparison to smooth flows. We do not possess a higher resolution simulation with tracer particles, however we verify in the right panel of Fig.~\ref{app2} that halo Au L8 simulated with 8 times poorer mass resolution shows very similar results to the fiducial simulation (left panel). Nevertheless, a larger simulation sample of very high resolution simulations (particularly at the disc-halo interface) is required to statistically quantify whether a highly resolved CGM affects the results presented here. Such a suite may become available in the near future, as numerical algorithms and computer power increase.

\begin{table}
\caption{Table showing the fraction of material locked up in stars at $z=0$ according to their classified origin, as in Fig.~\ref{forig}, for the fiducial Au L8, Au L8 run with 10 times as many tracer particles per cell, and Au L8 simulated with 8 times poorer mass resolution.}
\label{tab}
\centering
\begin{tabular}{c c c c}
\hline
Acc. type & fiducial $N_{\rm tr}$ & $10\times N_{\rm tr}$ & $8 \times$ lower res.\\
\hline
Star acc. & 0.09 & 0.07 & 0.10 \\
Sat Strip & 0.05 & 0.06 & 0.06\\
Sat Wind & 0.20 & 0.20 & 0.20\\
Sat ISM & 0.16 & 0.15 & 0.16\\
FF IGM & 0.42 & 0.43 & 0.38\\
IGM & 0.08 & 0.09 & 0.10\\
Tot. unrecycled & 0.10 & 0.11& 0.14\\
\hline     
\end{tabular}
\end{table}

\section{Conclusions}
\label{sec:conc}

We have studied the effects of galactic fountain flows on the evolution of angular momentum in a suite of magneto-hydrodynamical cosmological zoom-in simulations of the formation of Milky Way mass haloes. We conclude the following

\begin{itemize}
\item{} Approximately $90\%$ of material that ends up in stars at $z=0$ has been wind-recycled at least once. The median number of wind-recycling events across the simulation suite is 4.
\item{} Approximately half of the wind-recycled material is classified as gas that experienced its first wind-recycling event after its first smooth accretion from the IGM onto the central galaxy, subsequently entering into a galactic fountain (\ff). The other half originates from subhaloes, and was either ram-pressure stripped (\sts), ejected from the ISM of a subhalo before accretion (\sw), or accreted directly as ISM (\si) or star particles (\accs). 
\item{} Fountain flows occupy a characteristic zone around the central galaxy that extends to a few tens of kiloparsecs from the galactic centre, with median recycling timescales of about 500 Myr. Fountain flows generally act to increase the specific angular momentum of galaxies via the mixing of wind-recycled material with high angular momentum gas in the inner CGM/hot corona, which falls back onto the galaxy with higher angular momentum. 
\item{} The ability of galactic fountains to increase their angular momentum is sensitive to minor mergers and their orbital orientation. Gas-rich minor mergers on prograde orbits align the disc spin axis with that of the inner CGM and increase the specific angular momentum of surrounding gas, enabling the fountain to extract angular momentum from the inner CGM. Retrograde mergers produce the opposite effect, and act to lower the specific angular momentum of the disc. 
\item{} Bar-driven secular evolution can lead to a decrease in stellar specific angular momentum even in discs surrounded by a galactic fountain of higher specific angular momentum. 
\item{} Systems in which the galactic fountains gain the most angular momentum follow tracks parallel to, or steeper than, the disc sequence of the stellar mass-angular momentum relation \citep{FR18}. Retrograde minor mergers, bar-driven secular evolution and violent major mergers drive galaxies toward the bulge sequence relation. The interplay between these physical effects may explain why the trend and scatter of galaxies in this plane evolve little with redshift. 
\item{} Fountain flows act to flatten the radial metallicity gradient and narrow the metallicity dispersion of gas and newborn stars by moving metal-enriched gas to the outer disc via the CGM, where it is mixed efficiently.
\end{itemize}

In this paper, we studied the wind-recycling/fountain flow history of material found in the stars of Milky Way-mass galaxies. Future investigations into the evolutionary impact of wind-recycling in lower mass galaxies would be interesting, particularly to compare with other galaxy formation models \citep[e.g.][]{CDG16,AAF17}. Moreover, future simulations of much higher resolution will be extremely useful to analyze wind-recycling, fountain flows and gas mixing in finer detail.

\section*{Acknowledgements}
RG thanks Michael Fall for very useful discussions and comments on an earlier version of the manuscript, and the referee for providing a very constructive and useful report that led to the improvement of this paper. FvdV was supported by the Deutsche Forschungsgemeinschaft through project SP 709/5-1. JZ acknowledges funding from the European Research Council (ERC) under the European Union’s Horizon 2020 research and innovation programme (grant agreement No. 679145, project `COSMO-SIMS’). FAG acknowledges financial support from CONICYT through the project FONDECYT Regular Nr. 1181264, and funding from the Max Planck Society through a Partner Group grant. FM is supported through the Program ``Rita Levi Montalcini'' of the Italian MIUR.

\bibliographystyle{mnras}
\bibliography{mnras_template.bbl}

\bsp	
\label{lastpage}
\end{document}